\documentclass[aps,prb,amsmath,amssymb,twocolumn,superscriptaddress,showpacs,floatfix]{revtex4-1}
\usepackage{ifpdf}
 \newif\ifpdf
\ifx\pdfoutput\undefined
   \pdffalse
\else
   \pdfoutput=1
   \pdftrue
\fi
\ifpdf
   \usepackage{graphicx}
   \usepackage{epstopdf}
\else
   \usepackage{graphicx}
\fi

\usepackage[unicode=true,pdftex]{hyperref}\usepackage{srctex}

\usepackage{bm}
\usepackage{epsfig}
\usepackage[usenames]{color}
\usepackage{soul}
\usepackage{graphicx}
\usepackage{hyperref}
\usepackage[normalem]{ulem}
\usepackage{color}

\newcommand{\Cs}{\mathit C_{\scriptscriptstyle{\Sigma}}}
\newcommand{\Rs}{\mathit R_{\scriptscriptstyle{\Sigma}}}

\DeclareMathOperator{\ach}{arccosh}
\newcommand{\tP}{\tau_{\scriptscriptstyle{P}}}
\newcommand{\Ee}{\boldsymbol{\mathcal E}}
\newcommand{\te}{\tau_{\scriptscriptstyle{\mathcal E}}}

\newcommand{\Cso}{\mathit C^{(0)}_{\scriptscriptstyle{\Sigma}}}
\newcommand{\Eco}{E_c^{(0)}}

\begin{document}

\title{Single electron tunneling with ``slow'' insulators}

\author{S.~A.~Fedorov}
 \affiliation{Department of Theoretical Physics, Moscow Institute of Physics and Technology, Moscow 141700, Russia}
 \affiliation{P.N. Lebedev Physical Institute of the Russian Academy of Sciences, Moscow 119991, Russia}
\author{N.~M.~Chtchelkatchev}
\affiliation{Department of Physics and Astronomy, California State University Northridge, Northridge, CA 91330, USA}
\affiliation{Department of Theoretical Physics, Moscow Institute of Physics and Technology, Moscow 141700, Russia}
\affiliation{Institute for High Pressure Physics, Russian Academy of Science, Troitsk 142190, Russia}
\affiliation{L.D. Landau Institute for Theoretical Physics, Russian Academy of Sciences,117940 Moscow, Russia}
\author{O.~G.~Udalov}
\affiliation{Department of Physics and Astronomy, California State University Northridge, Northridge, CA 91330, USA}
\affiliation{Institute for Physics of Microstructures, Russian Academy of Science, Nizhny Novgorod, 603950, Russia}
\author{I.~S.~Beloborodov}
\affiliation{Department of Physics and Astronomy, California State University Northridge, Northridge, CA 91330, USA}

\date{\today}

\begin{abstract}
Usual paradigm in the theory of electron transport is related to the fact that the dielectric permittivity of
the insulator is assumed to be constant, no time dispersion. We take into account the ``slow'' polarization dynamics of the dielectric layers in the tunnel barriers in the fluctuating electric fields induced by single-electron tunneling events and study transport in the single electron transistor (SET). Here ``slow'' dielectric implies slow compared to the characteristic time scales of the SET charging-discharging effects.  We show that for strong enough polarizability, such that the induced charge on the island is comparable with the elementary charge, the transport properties of the SET substantially deviate from the known results of transport theory of SET. In particular,  the coulomb blockade is more pronounced at finite temperature, the conductance peaks change their shape and the current-voltage characteristics show the memory-effect (hysteresis). However, in contrast to SETs with ferroelectric tunnel junctions,~\cite{RefOurPRB,RefOurPRBSubm} here the periodicity of the conductance in the gate voltage is not broken, instead the period strongly depends on the polarizability of the gate-dielectric. We uncover the fine structure of the hysteresis-effect where the ``large'' hysteresis loop may include a number of ``smaller'' loops. Also we predict the memory effect in the current-voltage characteristics
$I(V)$, with $I(V)\neq -I(-V)$.
\end{abstract}

\pacs{77.80.-e,72.80.Tm,77.84.Lf}
\maketitle

\section{Introduction}

\begin{figure}[h]
  \centering
  \includegraphics[width=0.7\columnwidth]{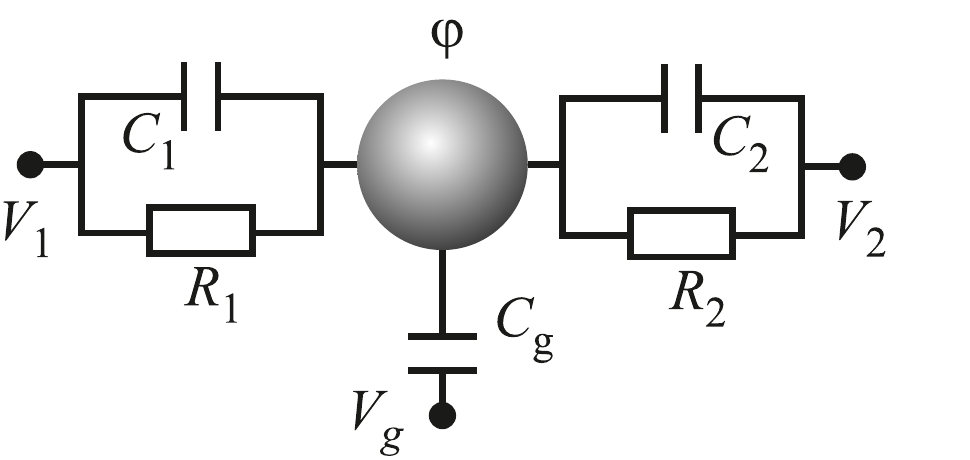}\\
  \caption{(Color online)   The equivalent scheme of single electron transistor (SET).}\label{fig_device}
\end{figure}

The single electron transistor (SET) is one of the most studied nanosystem,~\cite{RefOurPRB,RefOurPRBSubm,averin1991single,averin1991theory,devoret1992single,wasshuber2001computational} This is the
simplest device where strong electron correlations and quantum nature of electron can be directly observed.  It consists of two electrodes known as the drain and the source, connected through tunnel junctions to one common electrode with a low self-capacitance, known as the island. The electrical potential of the island can be tuned by a third electrode, known as the gate, capacitively coupled to the island, see Fig.~\ref{fig_device}.

For decades there was a paradigm in the theory of  electron transport at the nanoscale
related to the fact that the dielectric permittivity of nanojunctions was assumed to be constant, without any time dispersion.~\cite{arthur1954hippel,thoen1999handbook,ye2008handbook,Poplavko2009} However, this paradigm is not always true. A number of physical processes contribute to the polarization of dielectrics. Some of them are fast and some are slow compared to the time scales of electric field change in the nanojunctions. Recently, there was a progress in the development of new types of dielectric materials with strong and at the same time very slow response to the external electric field.~\cite{Dorogi1995PRB,Luo2006Chin} The SET is a perfect device where this physics can be studied. This is related to the fact that the charging-discharging effects in the SET
are controllable and have well-defined time scales.

The Coulomb blockade suppresses the electron transport except for values of the gate voltage where electrons
sequentially tunnel one by one through SET from source to drain.
The electric field in the tunnel junctions is changing in time while electrons tunnel through the island.
The dielectric layers in the tunnel junctions are polarized at finite electric field.
The usual assumption in the theory of SET is related to the fact
that the polarization of any dielectric layer in the tunnel barrier follows the electric field in time: $\mathbf P(t)=\alpha \, \boldsymbol{\mathcal E}(t)$, where the constant $\alpha$ is the dielectric permittivity of the dielectric layer.
It follows from the last expression that the capacitance $C$ of any tunnel junction in the SET is related to the geometric
capacitance $C^{(0)}$ as $C = \epsilon \, C^{(0)}$, where $\epsilon = (1+4\pi \alpha)$. And this is the only place where
the polarization appears in the theory of SET. However, these relations have limited applicability.
In general, the polarization of the dielectric is nonlocal in time: $\mathbf{P}(t)=\int_{-\infty}^t \chi(t-\tau)\Ee(\tau)d\tau$, where $\chi(t)$ is the dynamical electric permittivity. [Here we assume the linear response regime.] The time dependence of function $\chi(t)$ implies that tuning of dielectric polarization $\mathbf{P}(t)$ by an electric field can not be done arbitrary fast. This is happening, for example in dielectric materials with polarization being due to shift of heavy and inert ions.

The response of polarization $\mathbf{P}(t)$ to the external field is characterised by the
time-scale $\tP$, the decay time of function $\chi(t)$.
The second characteristic time-scale in the problem:
the time of the electric field correlation, $\te$.
For $\tP \ll \te$  the polarization has the form $\mathbf{P}(t)\approx \alpha\Ee(t)$,
where $\alpha=\int_{-\infty}^\infty\chi(\tau)d\tau$. In the opposite case, $\te\ll\tP$, the polarization $\mathbf P(t)$
does not follow the electric field $\Ee(t)$ instantaneously and it has the form
\begin{gather}\label{eqPav}
\mathbf P(t)\approx \alpha\,\langle\Ee\rangle,
\end{gather}
where $\langle\Ee\rangle$ is the electric field averaged over the time scale $\tP$. It follows from Eq.~\eqref{eqPav}
that the simple relation for capacitance, $C= \epsilon \, C^{(0)}$, is not valid at shorter times. Therefore
the theory of single-electron tunneling in the SET
should be modified and this is the main goal of our paper.

The characteristic time of charge relaxation in the SET is $\te=\Rs\Cs$, where $\Rs$ is of the order of the bare tunnel resistance of the left and right tunnel junctions and $\Cs$ is the sum of all the capacitances, see Fig.~\ref{fig_device}. The time scale $\te$ is
in the range of dozens of nano- to picoseconds depending on  the system geometry and materials. The switching
time of a dielectric material, $\tP$, is in the range of seconds to femto-seconds depending on
the material and the particular physical process behind the polarization phenomena.

Therefore the regime of ``slow'' insulator,  $\te\ll\tP$, is very important for SET-devices.However,
there is paradigm that the existing theories with $\tP \ll \te$ satisfactory explain most experiments with SETs.
What is the justification for new theory?
The answer is simple: the effects discussed in this paper are especially pronounced in SETs
when on average the polarization of a dielectric tunnel junction in the SET is strong enough meaning
that the charge induced on the grain by the polarized dielectric is of the order of the electron charge.
This condition can be reached for large enough dielectric permittivity $\epsilon$ only. How large we will discuss below.

Recently we have found a number of transport effects in the SET with slow
ferroelectric in the capacitors, see Refs.~\onlinecite{RefOurPRB,RefOurPRBSubm}.
In particular, we investigated the memory effect in this SET. Here we uncover new physical phenomena and
report about the memory-effect (hysteresis) where conductance periodicity in the gate voltage is not broken.
Instead, the period strongly depends on the polarizability of the gate-dielectric due to
the linear dependence of the polarization on the external field in the dielectric. Also, we
uncover the unusual fine structure of the hysteresis-effect, where ``large'' hysteresis loop may include
a number of ``smaller'' loops.  We predict that the memory effect exists in the current-voltage characteristics, meaning
that $I(V)\neq -I(-V)$ for a given memory branch even at $V_g=0$. The last two effects may exist in the ferroelectric SET, however
non of them have been found before.

The paper is organized as follows. In Sec.~\ref{SecI} we discuss the general properties of SET with slow
dielectric and the methods for investigation of transport properties.
In Sec.~\ref{SecII} we investigate the SET with slow dielectric located in the gate electrode at zero bias voltage, $V_2-V_1$.
In Sec.~\ref{SecIV} we consider the case with slow dielectric in the left and right tunnel barriers of the SET
and uncover the memory effect in the current-voltage characteristics, $I(V)$. Finally, in Sec.~\ref{SecDisc}
we discuss the validity of our approach and the requirements for slow dielectric materials which are
necessary to observe the effects predicted in this paper.
In the same section we show that the Coulomb blockade in SET with slow dielectrics is less affected by temperature.

\section{Electron transport through set with slow tunnel barriers\label{SecI}}

Here we consider the theory of SET with slow barriers. In the following it is convenient to distinguish between the geometrical junction capacitances $C_i^{(0)}$ and the low-frequency capacitances $C_i$ that include the slow dielectric response. The difference between them, aside from the unimportant geometrical factor, is
\begin{equation}
\Delta C_i=C_i-C_i^{(0)}=\alpha_i S_i/ d_i,
\end{equation}
where $\alpha_i$ is the dielectric polarizability of the i-th junction ($i=1,2,g$), $S_i$ --- the junction surface area and $d_i$ --- the effective electrode-island distance.

We assume that the electrodes are biased with the voltages $V_1=-V/2$, $V_2=V/2$ and $V_g$.
The grain potential $\phi(n)$ at a given number of excess electrons $n$ can be found balancing the induced charges:
\begin{gather}\label{EqCharge0}
n\,e=\sum_iC_{i}^{(0)}(\phi(n)-V_i)+\sum_i \Delta C_i(\langle\phi\rangle-V_i),
\\ \label{EqCharge1}
\langle\phi\rangle=\sum_{n=-\infty}^\infty p_n \phi(n),
\end{gather}
where $p_n$ is the probability to find $n$ excess charges on the grain. Two terms originate in~\eqref{EqCharge0} because we distinguish the electric field produced by the capacitance $C_{i}^{(0)}$ and the contribution due to polarized dielectric with slow response. So the terms proportional to the coefficient $\Delta C_i$ in Eq.~(\ref{EqCharge1}) can be considered
as charges induced on the grain by the polarized dielectric layers that are constant in tunneling events.
\begin{figure}[b]
  \centering
  \includegraphics[width=0.85\columnwidth]{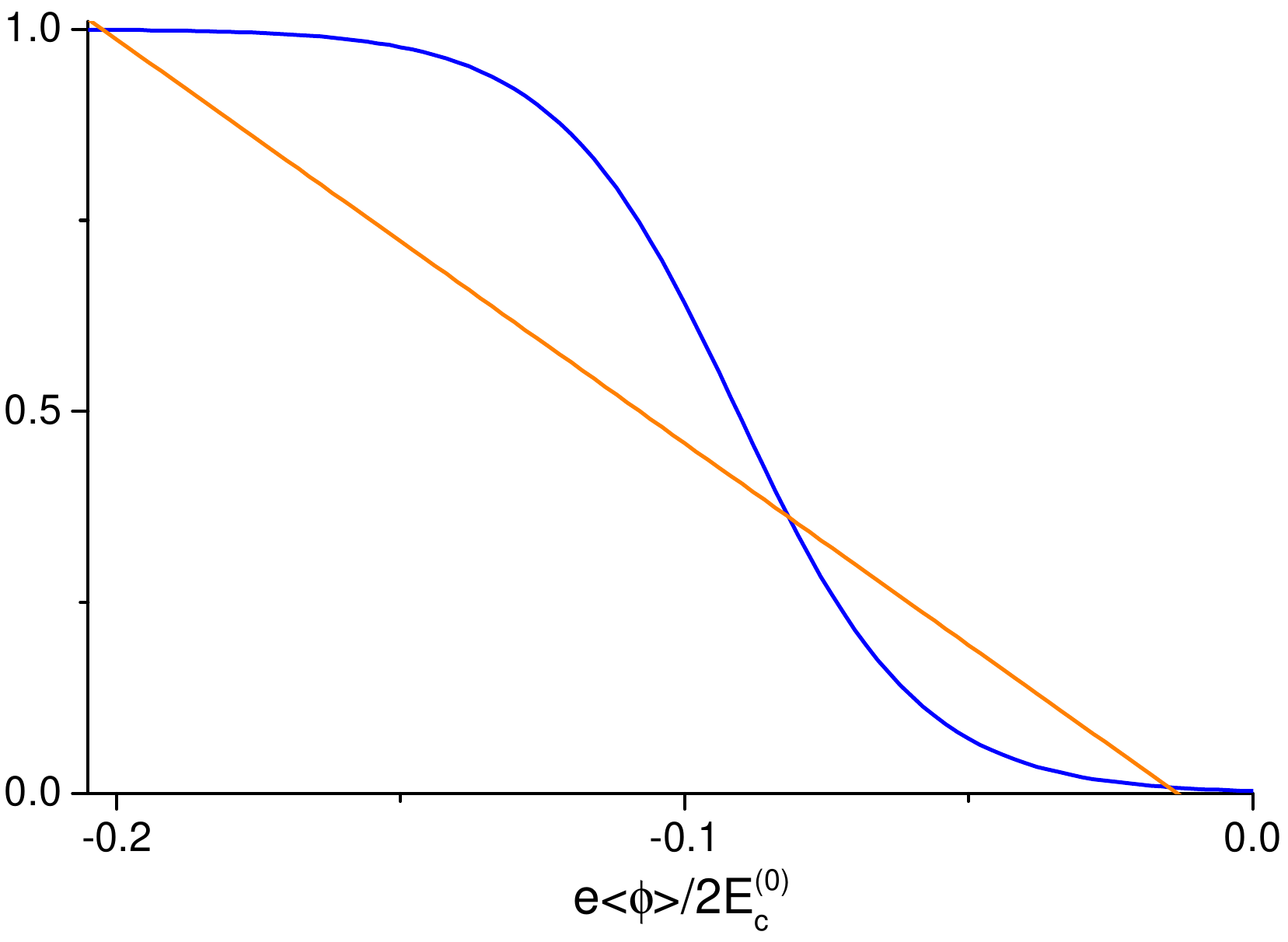}\\
  \caption{(Color online) Graphical solution of Eq.~\eqref{eqnf} showing three possible
  solutions for an average grain potential $\langle\phi\rangle$ at a given gate
  voltage $V_g$. Parameters are: $Q=-0.07|e|$, $C^{(0)}_g=0.5\Cso$, $\Delta C_g/\Cso=0.5$, and $T=0.4\,E_c^{(0)}$.
  The three distinct solutions for $\langle\phi\rangle$ at a given $Q_0$ correspond to
  the memory effect instability. \label{figaprox}}
\end{figure}

The probability distribution $p_n$ in the steady state can be found using the detailed balance equation~\cite{averin1991single,averin1991theory,devoret1992single,wasshuber2001computational}
\begin{equation} \label{eq:ballance}
    p_n\Gamma^{n\to n+1}=p_{n+1}\Gamma^{n+1\to n},
\end{equation}
where the rate $\Gamma^{n\to n+1}[V_i,n,\langle\phi\rangle]$ describes the change of grain charge from $n$ to $n+1$ electrons.
The electric current has the form
\begin{equation}
    I = e\sum_{n=-\infty}^\infty p_n \left[ \Gamma_{s}^{n\to n-1}-\Gamma_{s}^{n\to n+1} \right]\,.
\label{eq:I}
\end{equation}
Here the lower index of $\Gamma$ refers to the tunneling rate corresponding to the particular tunnel junction, $s=1$ or $2$ and
the rate $\Gamma$ in Eq.~\eqref{eq:ballance} is equal to $ \Gamma_{1}+ \Gamma_{2}$. Solving Eqs.~\eqref{EqCharge0}-\eqref{eq:ballance} self-consistently we find the current-voltage characteristics of the SET
using Eq.~\eqref{eq:I}.

We use the ``orthodox'' theory to calculate the Coulomb-blockade peaks in the differential conductance of the SET.
The calculation of $\Gamma$-rates requires the knowledge of the difference in the electrostatic energies when the number of excess charges on the grain differ by one elementary charge: $\Delta U^\pm_n=U(n\pm1)-U(n)$.
If the polarization in dielectric layers on electron jumps follow $\phi$ adiabatically, $P_i = \alpha_i (\phi-V_i)/d_i$,
we have $\Delta U^\pm_n=E_c(1\pm 2n)$,
where $E_c=e^2/2\Cs$ with all the capacitances $\Cs=\sum_i C_i$ being properly renormalized,
$C_i=C_i^{(0)}(1+4\pi\alpha_i)$. However, for slow dielectric layers the polarization
$P_i=\alpha_i (\langle\phi\rangle-V_i)/d_i$ stays constant during the tunneling, and
for the energy difference we find (see App.~\ref{Ap2})
\begin{gather}
  \Delta U^\pm_n=\Eco\,(1\pm2n\mp2\sum\nolimits_iP_iS_i/e),
\label{eqDeltaUSlow}
\end{gather}
where $\Eco=e^2/2\Cso$, $\Cso=\sum_i C_i^{(0)}$ and $P_i S_i=\Delta C_i (\langle\phi\rangle-V_i)$.

The work done by the leads and the gate to transfer an electron to/from the grain remains the same as in
the ``orthodox'' theory except for the fact that only the geometrical capacitances $C_i^{(0)}$ should be taken into account. This implies that for temperature $T\to 0$ the effective ground state free energy is defined as
\begin{equation}
F_0=E_c^{(0)}\min_n(n-Q'/e)^2,
\end{equation}
where the effective gate-induced charge $Q'$ is
\begin{equation}\label{eqQg}
  Q'=-C_g^{(0)} V_g+\sum\nolimits_i \Delta C_i\, (\langle\phi\rangle-V_i).
\end{equation}

\begin{figure}[t]
  \centering
  \includegraphics[width=0.99\columnwidth]{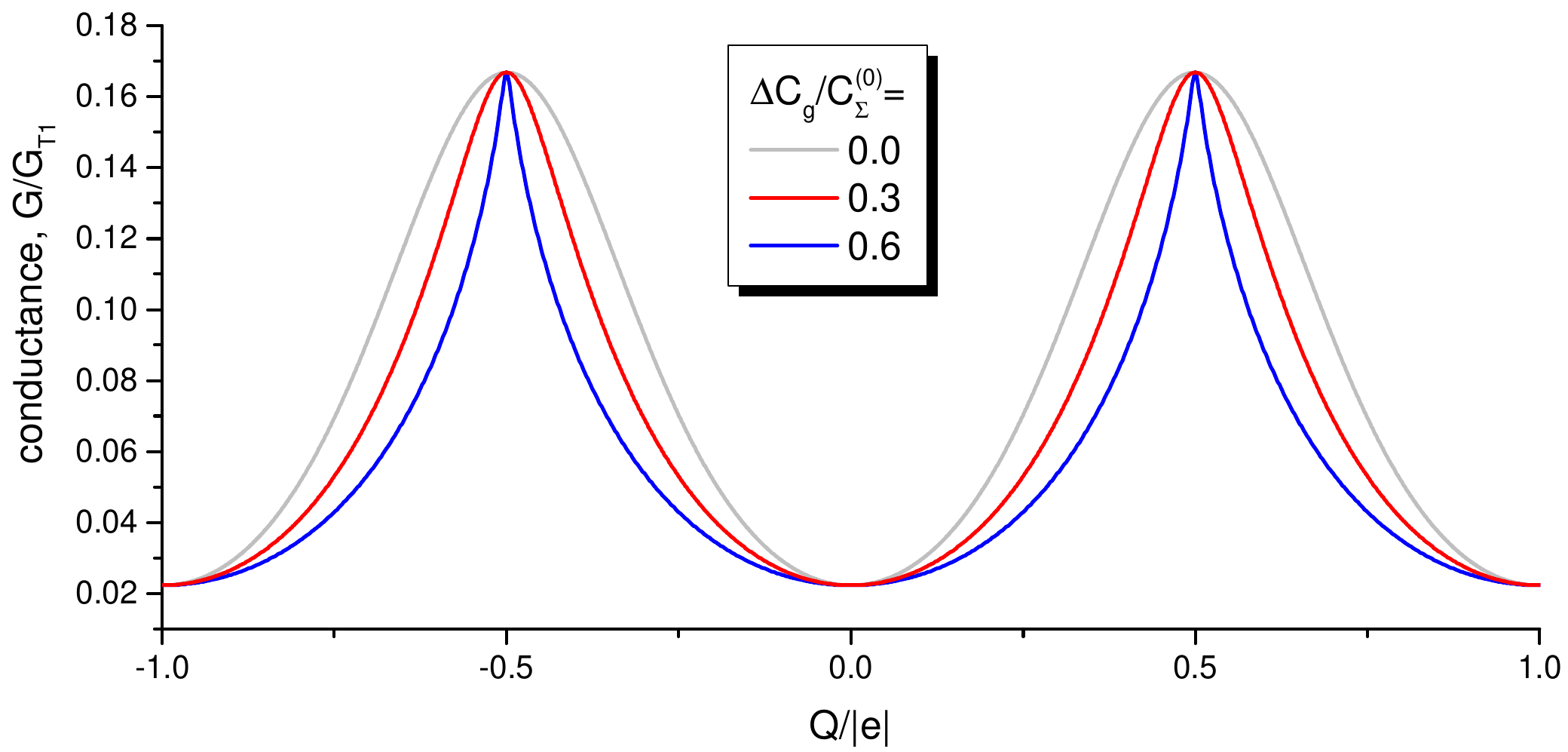}\\
  \caption{(Color online) Conductance peaks for $\Delta C_g/\Cso=0,\,0.3,\,0.6$. The ``unit'' of conductance $G_{T1}$ is the conductance of the first tunnel junction of the SET. Parameters are: capacitances $C^{(0)}_1=0.3\Cso$, $C^{(0)}_2=0.5\Cso$ and $C^{(0)}_g=0.2\Cso$ and temperature $T=0.2\Eco$. The slow dielectric in the gate capacitor modifies
  the shape of the conductance peaks but preserves the periodicity in parameter $Q$
  in contrast to the SET with ferroelectric in the gate capacitor.~\cite{RefOurPRBSubm}  }\label{figagsmall}
\end{figure}
Below we use the notation $Q=-C_g V_g$ for the traditional gate-induced charge. We show that although the effects of slow polarization are far from being a simple renormalization of capacitances $C_i^{(0)}\to C_i$, the conductance periodicity in $Q$ holds and maintains its period $|e|$ for any values of the parameters $\Delta C_i$.
\begin{figure}[t]
  \centering
  \includegraphics[width=0.99\columnwidth]{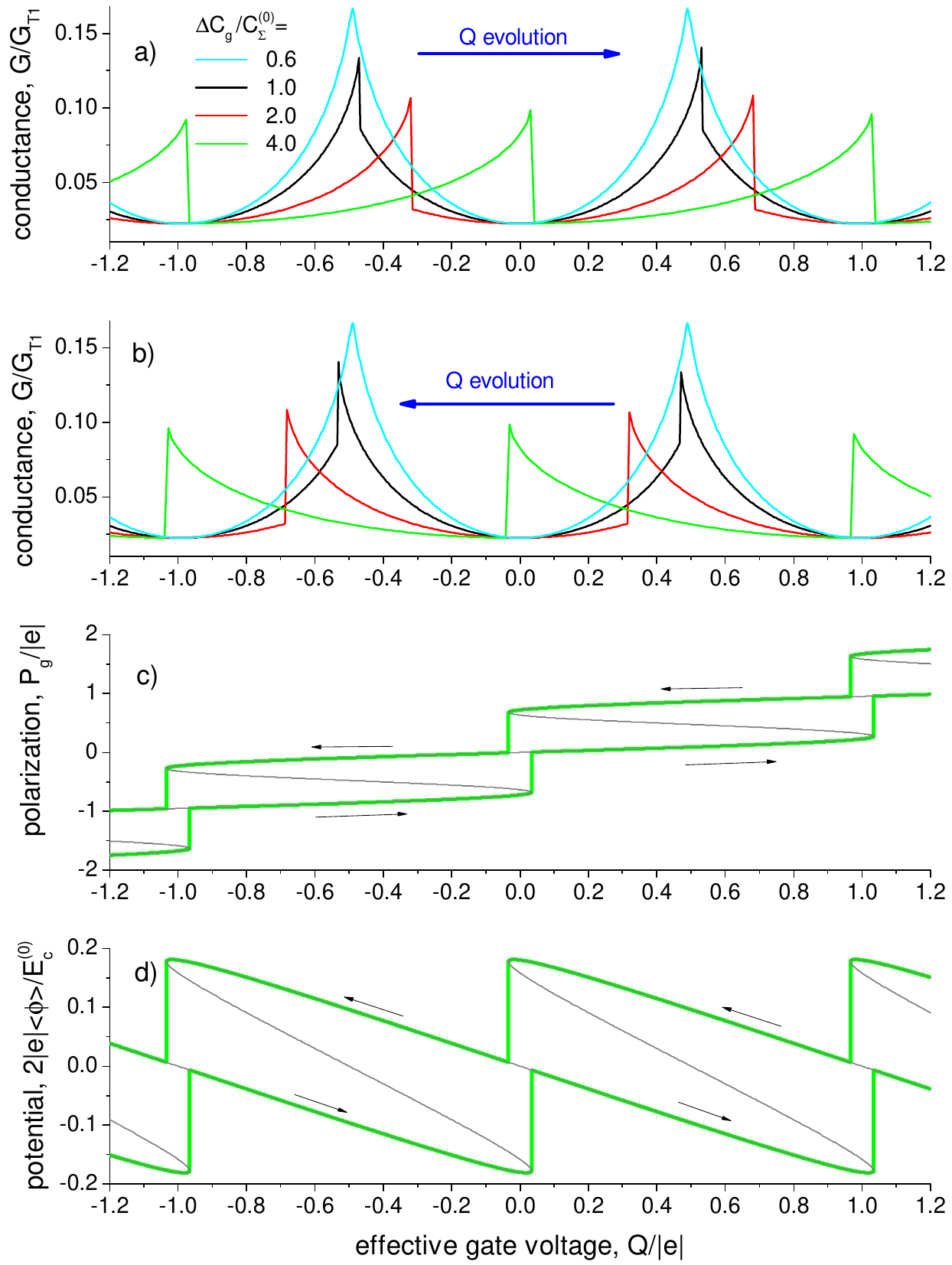}\\
  \caption{(Color online)  Memory effect instability in the SET with slow insulator in
  the gate capacitor. (a) and (b) the conductance branches corresponding to the increasing and decreasing parameter
  $Q$ and $\Delta C_g/\Cso=0.6,\,1,\,2,\,4$, (c) polarization and (d) the average grain potential
  (arrows show the direction of $Q$ evolution for a given branch) for $\Delta C_g/\Cso=4$.
  Grey lines show stable and unstable branches of polarization and the average potential.
  Parameters are: capacitances $C^{(0)}_1=0.3\Cso$, $C^{(0)}_2=0.5\Cso$ and $C^{(0)}_g=0.2\Cso$ and temperature $T=0.2\Eco$ as in Fig.~\ref{figagsmall}. }\label{fighist}
\end{figure}
\begin{figure*}[t]
  \centering
  \includegraphics[width=\textwidth]{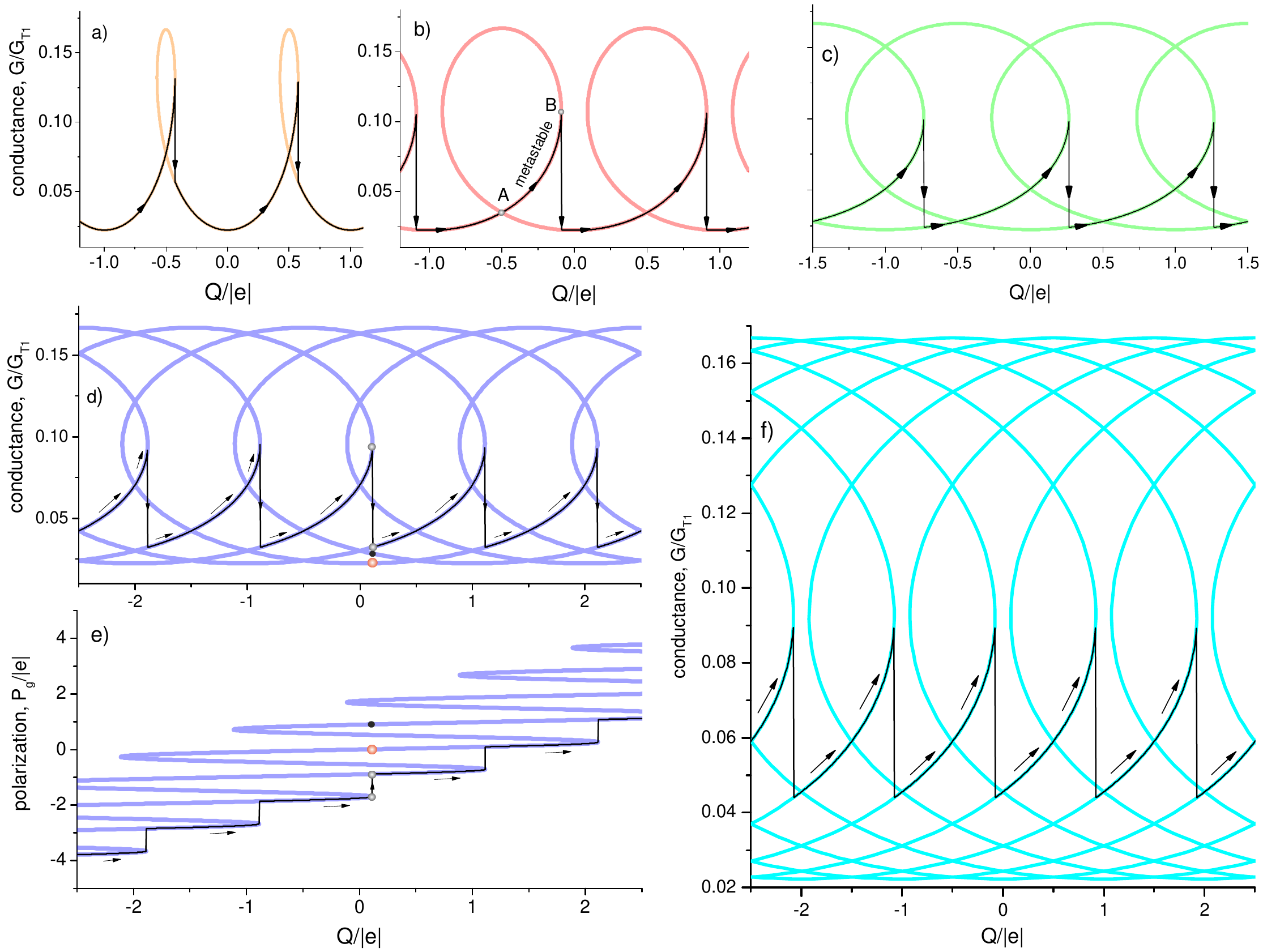}\\
  \caption{(Color online) Memory effect: Plots (a)-(d) and (f) show the conductance for $\Delta C_g/\Cso=1.3,\,3.3,\,5.3,\,10,\,20$  for
  stable and unstable branches of Eq.\eqref{EqCharge1} for the average grain potential. Plot (e) shows the polarization for $\Delta C_g/\Cso=10$.
  Arrows indicate the position of hysteresis jumps for particular branch with increasing $Q$. All plots are shown at fixed temperature  $T=0.2\Eco$.}\label{figbig}
\end{figure*}

The detailed balance equation~\eqref{eq:ballance} can be solved
analytically for the set of voltages $V_g$ near the ``degeneracy points'',
where the ground state energy of the SET changes from $n$ to $n\pm 1$ excess charges.
The last condition requires the effective charge $Q'$ to be close to $e(n+1/2)$. In this case the only two
probabilities $p_n$  are finite while the other probabilities
are exponentially suppressed by the factor $e^{-E_c^{(0)}/T}$.
In order to illustrate the origin of the memory effect,
we will focus on the degeneracy point between $n=0$ and $n=1$
at $V_{1,2}=0$. Using Eqs.~\eqref{EqCharge0}-\eqref{EqCharge1} we find for the average potential
$\langle\phi\rangle$
\begin{gather}\label{eqnf}
  n_F[(1-2Q'/e)\Eco]=e\langle\phi\rangle/2\Eco+Q'/e,
\end{gather}
where $n_F$ is the Fermi-function. Equation~(\ref{eqnf}) has one or three solutions for a given
gate voltage $Q$.
The latter case is shown in Fig.~\ref{figaprox}. The presence of three distinct
solutions for the average potential $\langle\phi\rangle$ at a given parameter $Q$ indicates
the memory effect instability. Using the graphical solution of Eq.~\eqref{eqnf} we estimate
the criteria for
the memory effect instability, $\sum_i \Delta C_i/\Cso \gtrsim2 T/\Eco$.
This criterion corresponds to the critical value of $\Delta C_{\Sigma}$
when the memory effect just appears, see Eq.~\eqref{alphacn1} below for the exact expression.

\section{SET with slow insulator in the gate capacitor\label{SecII}}

\subsection{Numerical study of electron transport through SET}

Here we study electron transport through SET numerically.
We consider the SET with slow dielectric layer in the gate capacitor.
This set-up is the most favourable for experiment
since in this case there is no electron tunneling through the gate electrode and it can be arbitrary
thick to allow a wide choice of dielectric materials. Moreover, as we will show
in the following Sec.~\ref{SecIV}, at $V=0$  by considering the gate capacitor we still preserve all the qualitative effects introduced by slow dielectrics in a general case.
\par
Thus, for a time, we assume that the only non-zero $\Delta C$ is $\Delta C_g$.

For $\Delta C_g = 0$ the conductance is a periodic function of the effective gate voltage $Q$, see the
gray curve in Fig.~\ref{figagsmall}. The conductance peaks are well fitted by the orthodox theory
where near the peak maximum the conductance is
\begin{gather}\label{eqG0}
  G^{(0)}(\delta Q^{(0)})\approx \frac{e\,\delta  Q^{(0)}/\Cso T}{2(R_1+R_2)\sinh(e\delta Q^{(0)}/\Cso T)}.
\end{gather}
Here $\delta Q^{(0)}/e = \min_k[-C_g^{(0)}V_g/e - (2k+1)/2] \ll 1$.

At finite but small $\Delta C_g$, when the induced charge on the island due to polarization is
smaller than the elementary charge, the conductance peaks change their shape,
but preserve their amplitude and position (see Fig.~\ref{figagsmall}).

The opposite case, with dielectric polarization being strong enough to induce the charge on the island of the order of the elementary charge or larger, is more interesting. In this case the conductance peaks show the hysteresis and their shape depends on the direction of $Q$-evolution, see Fig.~\ref{fighist}. The hysteresis appears for $\Delta C_g\gtrsim \Cso 2T/\Eco$ (see Eq.~\ref{alphac}).
Despite the memory effect the conductance remains periodic in the renormalized gate voltage $Q=-(C^{(0)}_g+\Delta C_g)V_g$ with the same period $|e|$ for any $\Delta C_g$. This behavior is in striking contrast to the SET with ferroelectric in the gate where due to the nonlinearity of polarization--electric field dependence the periodicity of conductance is broken, see Ref.~\onlinecite{RefOurPRBSubm}.
\begin{figure}[t]
  \centering
  \includegraphics[width=0.99\columnwidth]{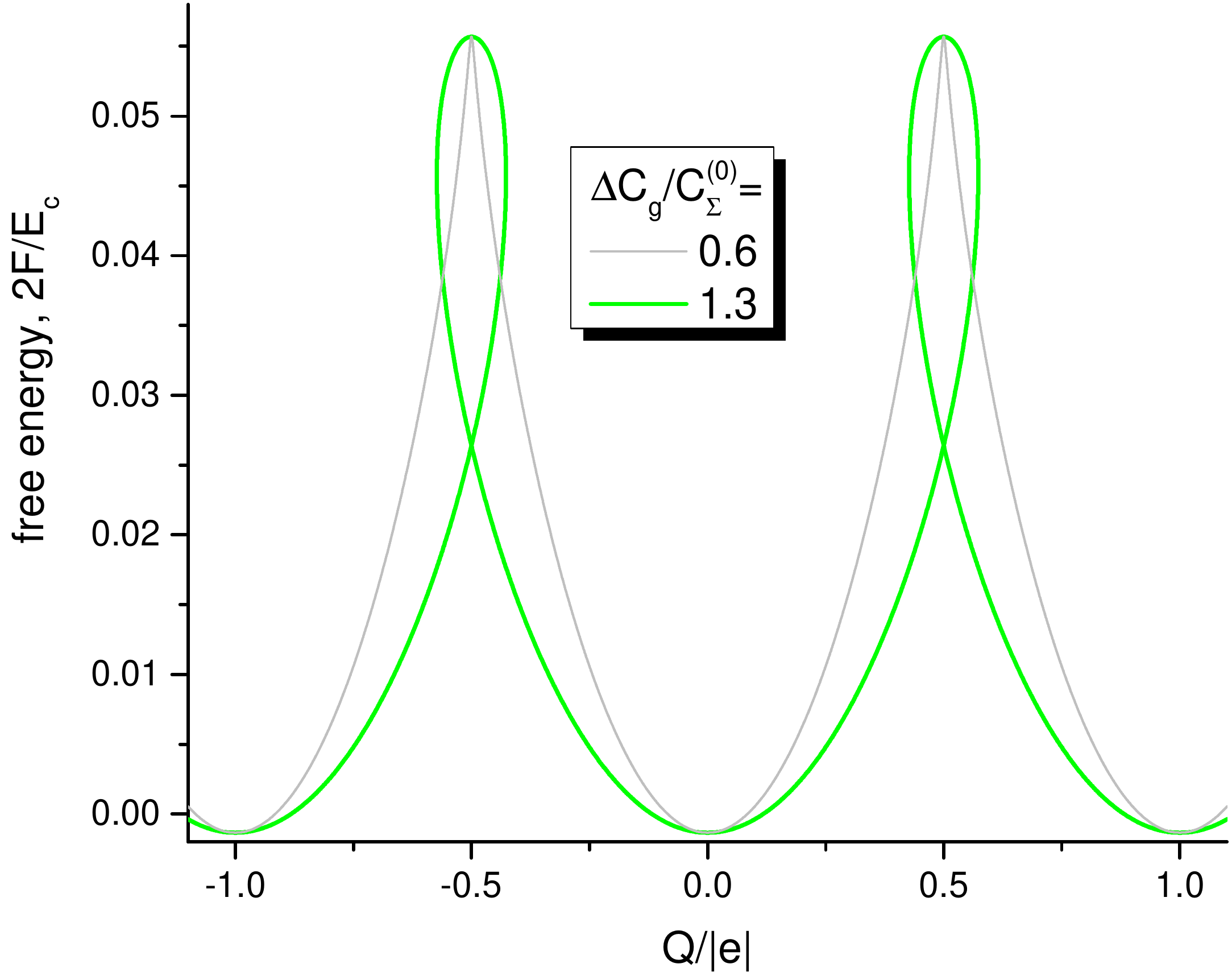}\\
  \caption{(Color online) Free energy in Eq.\eqref{eqF} for $\Delta C_g/\Cso=0.6,\,1.3$ and temperature
  $T=0.2\Eco$. The shape of free energy plot is similar to the conductance $G(Q)$ plot in
  Fig.~\ref{figbig}a. }\label{figF}
\end{figure}

Now we discuss the structure of the memory effect. Above the critical value of $\Delta C_g$ there are many brunch-solutions of the self-consistency equation for the average grain potential, Eq.~\eqref{EqCharge1}, for the given temperature,
bias and gate voltage. The question is - how to choose the right branch? Figure~\ref{figbig} provides an
answer to this question. According to the branching theory~\cite{VainbergTrenogin} the jumps occur at the ``branching points'' where the observable has an infinite derivative in parameter $Q$.
On the other hand, the branch should correspond to the minimum of some effective
energy functional. In our case (no bias) the role of the effective energy plays the free energy
\begin{gather}\label{eqF}
  F=-T\ln Z,\qquad Z=\sum_n \exp\left(-\frac{E_c^{(0)}\,(n-Q'/e)^2}{T}\right).
  \end{gather}
For zero temperature it reduces to the free energy $F_0$ discussed above.

The plots of the free energy have a similar dependence on the parameter $Q$ as the zero-bias conductance $G$.
To illustrate this point we show in Fig.~\ref{figF} the free energy for $\Delta C_g/\Cso=0.6,\,1.3$. Figure~\ref{figbig}b shows
that the conductance branch between points ``A'' and ``B'' is metastable: the free energy for
this curve is larger than the free energy for
branch below. However, during the adiabatically slow increase of parameter $Q$ the system does not switch to the lowest branch at point A,
instead it may go up to the metastable branch. The same applies to all other plots in Fig.~\ref{figF}.
The external perturbation can drive the system to outside of the metastable branch before the bifurcation point.
Usually the role of this ``perturbation'' plays the Langevin forces induced by the thermostat. In this case
the jumps occur randomly within the same region before the bifurcation point.
This scenario is typical for any hysteresis.

Intuitively one may suppose that if conductance ``jumps'' from one branch to another the final branch should have
the lowest possible free energy for the parameter $Q$ corresponding to the jump.
Indeed, this is the case in Figs.~\ref{figbig}(a)-(c). However, in
Figs.~\ref{figbig}(d) and (f) this rule is violated.
The system could jump, for example, to the point marked by the red-ball in Fig.~\ref{figbig}(d),
instead of finishing at the point marked by the grey-ball which has a larger free energy. However, this
energetically favourable transition is ``forbidden'':
while continiously changing the polarization in such a process the system would have to pass the energy barrier of approximately $\Eco/4$ (free energy maximum). Thus the higher order jumps (over the average charge difference)
are suppressed by the factor $\exp(-E_c^{(0)}/4T)$.

\subsection{The fine structure of the memory effect}
\begin{figure}[t]
  \centering
  \includegraphics[width=\columnwidth]{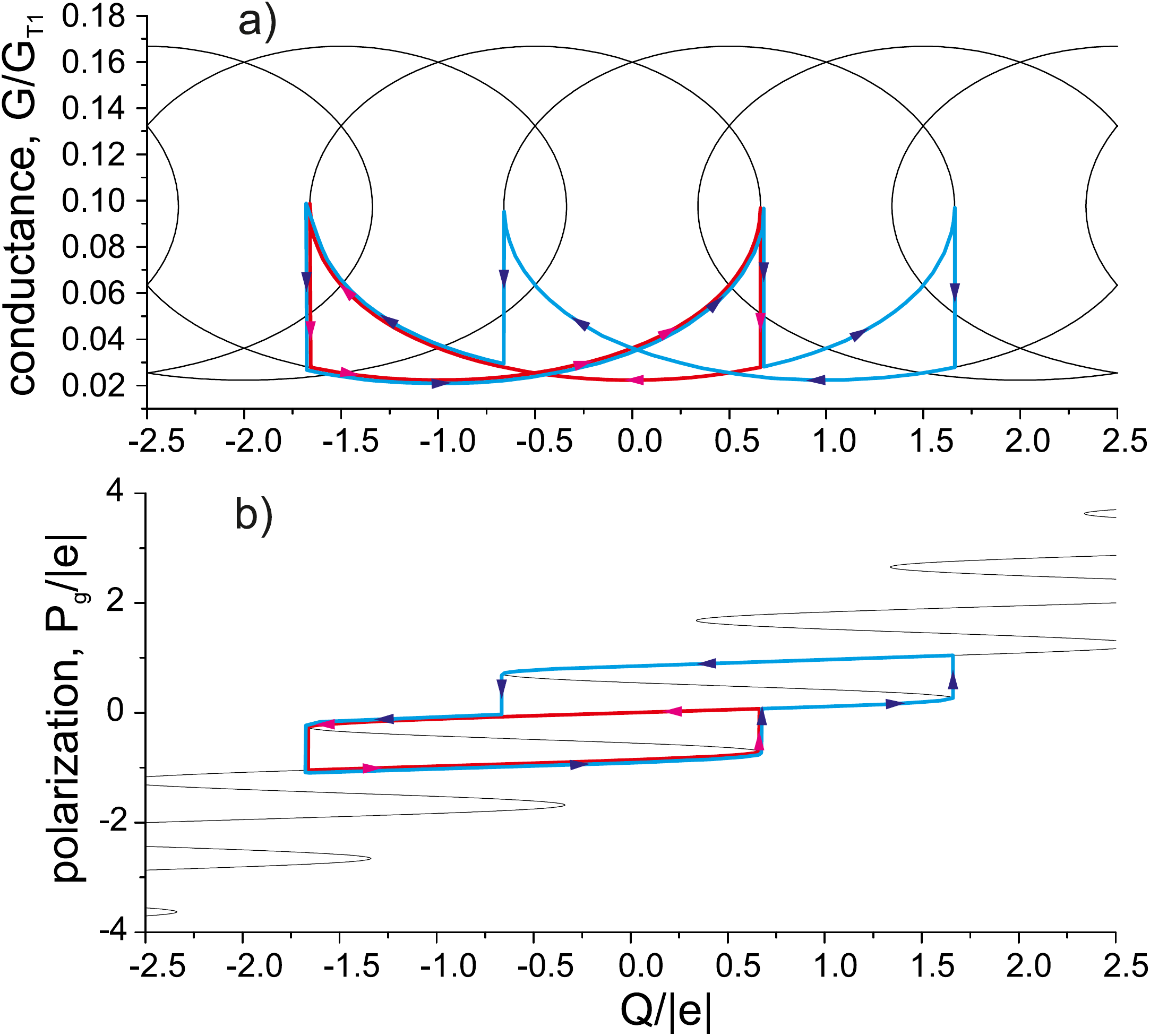}\\
  \caption{(Color online) Memory effect in  (a) the conductance and (b) the polarization of the gate-insulator. The red
  hysteresis loop corresponds to back and forth change of parameter
  $Q$ in the interval $(-2,1)$, while the blue curve corresponds to $(-2,2)$ interval.
  Parameters are: $\Delta C_g=7.5\Cso$, $T = 0.2\,E_c^{(0)}$, while $C_i^{(0)}$, $i=1,2,g$
  and $R_j$, $j=1,2$ similar to Fig.~\ref{figagsmall}. }\label{fighistsmall}
\end{figure}

Doing numerical studies of memory effect we assumed that parameter $Q$ increases (or decreases) monotonically
from minus to plus infinity (or vice-versa). However, for large enough parameter $\Delta C_g$,
when polarization induces more than one electron on the grain, the hysteresis loop depends on the interval where
the  parameter $Q$ changes. This is shown in Fig.~\ref{fighistsmall} with two possible hysteresis loops:
The red hysteresis loop corresponds to back and forth change of parameter $Q$ in the
interval $(-2,1)$ while the blue curve corresponds to the interval $(-2,2)$. In the second case the
larger hysteresis loop ``includes'' smaller loops. As a result, the understanding of memory effect at finite
intervals of parameter $Q$ evolution requires consideration of all branches of the SET observables such as
conductance and polarization.

\subsection{Analytical description of the conductance peaks and the memory effect}
\begin{figure*}[t]
  \centering
  \includegraphics[width=\textwidth]{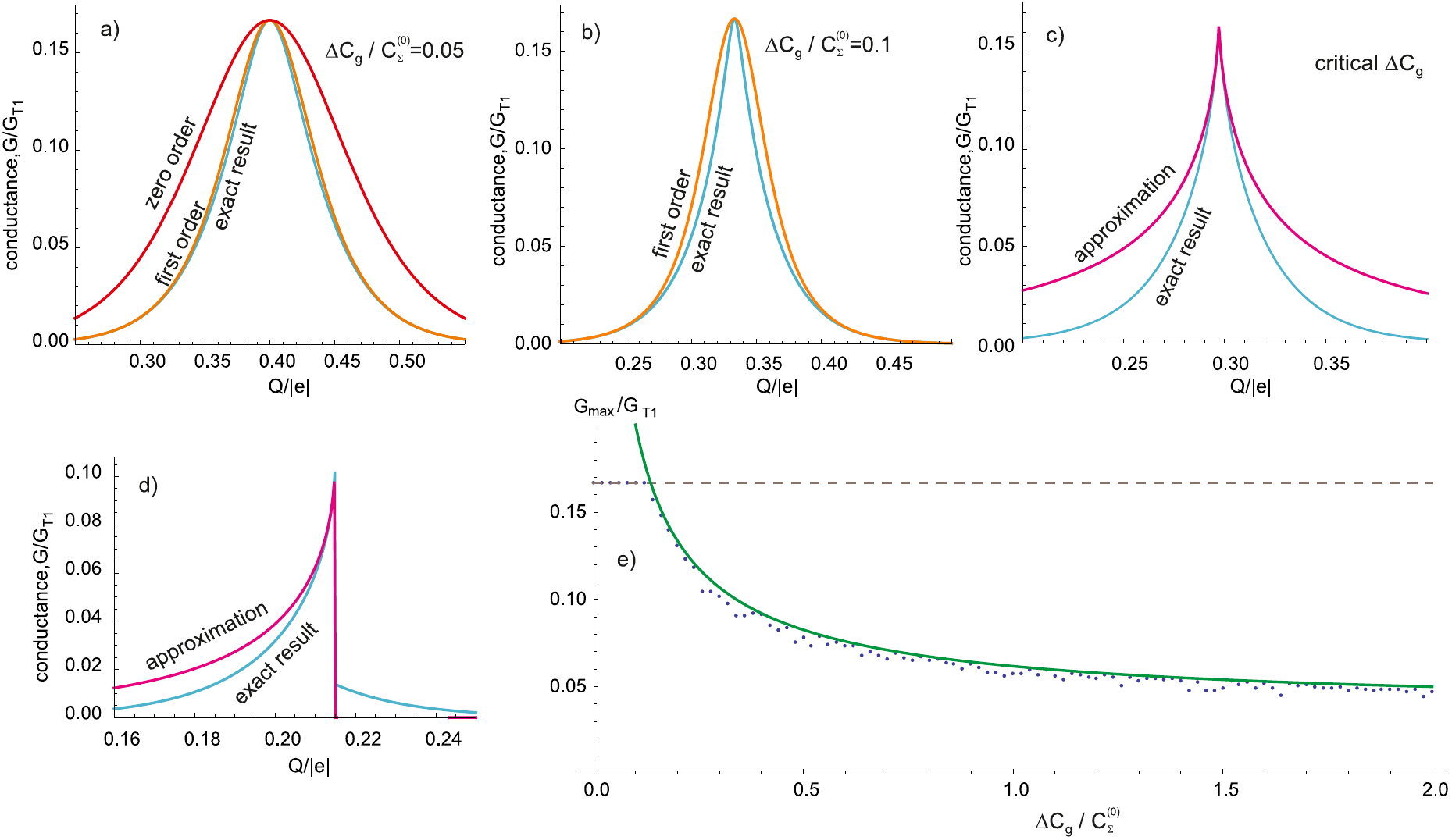}\\
  \caption{(Color online) (a) Numerical solution for conductance peak for $\Delta C_g/\Cso=0.05$  (blue line), orange
  line is the zero order solution from Eq.~(\ref{Q00ord}), and the red line is the first order solution from
   Eq.~(\ref{Q01ord}). (b) Numerical and analytical solutions of conductance for $\Delta C_g/\Cso=0.1$. The first order
  approximates well the conductance peak at small $\Delta C_g$. (c) Conductance at critical $\Delta C_g$, where
  hysteresis appears. (d) Conductance hysteresis. (e) Amplitude of conductance peak
  vs $\Delta C_g$. Points represent the numerical solution; the red curve is $G_{\rm max}=1/2(R_1+R_2)$; and
  the orange curve shows Eq.~\eqref{GmaxEq} for $G_{\rm max}$.
  Parameters are: $T = 0.06\,E_c^{(0)}$, $C^{(0)}_i$, $i=1,2,g$ and $R_j$, $j=1,2$ similar to
  Fig.~\ref{figagsmall}.}\label{figaproxn}
\end{figure*}

Here we present the analytical description of transport properties of SET.
At $V=0$ and within the two-state approximation the form of the conductance peaks $G(Q)$ can be
found using Eq.~\eqref{eqG0} with the proper substitution $Q^{(0)}\to Q'$, where $Q'$ is defined in Eq.~\eqref{eqQg}: with this
substitution we have for conductance $G(Q)=G^{(0)}(Q')$.
For average potential, generalizing Eq.~\eqref{eqnf}, we obtain
\begin{equation}\label{eqphi}
\langle\phi\rangle=\frac{e}{\Cso}\left(\frac{1}{2}\mathrm{tanh}\left(\frac{E_c^{(0)}}{T}\frac{\delta Q'}{e}\right)-\frac{\delta Q'}{e}\right),
\end{equation}
where $\delta Q'/e=\min_k( Q'/e-(k+1/2))$. Combining Eqs.~\eqref{eqphi} with \eqref{eqQg} we find,
\begin{equation}\label{sce2}
\delta Q'\frac{C_\Sigma}{\Cso}-\frac{e}{2}\frac{\Delta C_g}{\Cso}\mathrm{tanh}\left(\frac{E_c^{(0)}}{T}\frac{\delta Q'}{e}\right)=\delta Q,
\end{equation}
where $\delta Q=Q-(k+1/2)e$  is the deviation of
parameter $Q$, $k$ is the same as for $\delta Q'$ and $C_\Sigma=\Cso+\Delta C_g$.
It should be noted that the above equations are valid for any $\delta Q$ as long as  $\delta Q'\ll 1$.

\subsubsection{Small polarization}

Here we discuss the limit of small polarization, meaning that the induced charge on the island is small
compared to the elementary charge $e$.
Using the small parameter, $\Delta C_g/\Cso\ll 1$, we expand Eq.~\eqref{sce2} up to the second order
\begin{align}
 (\delta Q')_{0} &= \delta Q\frac{\Cso}{C_\Sigma}, \label{Q00ord}\\
 (\delta Q')_{1} &=(\delta Q')_{0}+\frac{e}{2}\frac{\Delta C_g}{C_\Sigma }\mathrm{tanh}\left(\frac{E_c^{(0)}}{T} \frac{(\delta Q')_{0}}{e}\right). \label{Q01ord}
\end{align}
The conductance now may be found by substituting $\delta Q^{(0)}$ with $(\delta Q')_{0,1}$ in Eq.~\ref{eqG0}.
\begin{equation}\label{eqGn}
G(\delta Q)=G^{(0)}(\delta Q')
\end{equation}

The numerical calculations in Fig.~\ref{figaproxn}(a) show that the first order
approximation, Eq.~\eqref{Q01ord}, well describes the peak shape for small parameter
$\Delta C_g/\Cso\approx 0.1$, while the zero order approximation is not sufficient.
We note that parameter $\Delta C_g/C_g^{(0)}$ and thus the renormalization of the conductance period over $V_g$
can be arbitrary in this approximation.

\subsubsection{Amplitude and form of the conductance peak in the hysteresis regime}

Solution of Eq.~\eqref{sce2} becomes ambiguous for large values of parameter $\Delta C_g$, where conductance $G(Q)$
acquires hysteresis. In this case the form of conductance peaks becomes nonsymmetric and
the conductance $G(Q)$ has a maximum at the branching (bifurcation) point corresponding to the
jump of the polarization. The bifurcation points in Eq.~\eqref{sce2} can be found as follows
\begin{equation}
\frac{d}{dQ'}\left(\delta Q'\frac{C_\Sigma}{\Cso}-\frac{e}{2}\frac{\Delta C_g }{\Cso}\mathrm{tanh}\left(\frac{E_c^{(0)}}{T}\frac{\delta Q'}{e}\right)\right)=0,
\end{equation}
that reduces to
\begin{equation}
\cosh^2\left(\frac{E_c^{(0)}}{T}\frac{(\delta Q')_{\rm max}}{e}\right)=\frac{E_c^{(0)}}{2T}\frac{\Delta C_g}{C_\Sigma}.
\label{maxPoint}
\end{equation}

The two solutions of Eq.~\eqref{maxPoint} correspond to the increasing and
decreasing evolution of parameter $Q$ (solutions with $\delta Q'<0$ and $\delta Q'>0$ respectively).
These two solutions result in mirror-reflected shapes for the peaks, so we focus only
on the decreasing parameter $Q$. For conductance maximum we find
\begin{equation}\label{GmaxEq}
G_{\rm max}=\frac{1}{2(R_1+R_2)}\frac{\ach\left(\sqrt{\frac{E_c^{(0)}}{2T}\frac{\Delta C_g}{C_\Sigma}}\right)}{\sqrt{\frac{E_c^{(0)}}{2T}\frac{\Delta C_g}{C_\Sigma}\left(\frac{E_c^{(0)}}{2T}\frac{\Delta C_g}{C_\Sigma}-1 \right)}}.
\end{equation}
The predicted conductance maximum amplitude variation
is shown in Fig.~\ref{figaproxn}. One can see that the curve breaks at critical value of parameter
$\Delta C_g$ indicating the start of the hysteresis regime.

We note that since within the scope of the two-state approximation and for $\Delta C_g$ above the critical value the  Eq.~\eqref{GmaxEq} gives exact maximum, its applicability depends only on temperature. At finite $\Delta C_g$ the conductance maximum does not exactly correspond to a degeneracy point $\delta Q'=0$, but still $\delta (Q')_{max}\ll1$ for $T\ll \Eco$. For example, for temperature
$T=0.06 \Eco$ and $\Delta C_g\to\infty$ we have $\delta (Q')_{\rm max}/|e|\approx 0.1 \ll 1$, meaning that our consideration is valid (see Fig.~\ref{figaproxn}).

Now we find the form of conductance peaks. Expanding Eq.~\eqref{sce2} up to the second order near $\delta (Q')_{\rm max}$
we obtain
\begin{equation}
A_0+A_2(\delta Q'-\delta (Q')_{\rm max})^2=\delta Q,
\label{sce6}
\end{equation}
where
\begin{multline}
A_0=\frac{eT}{E_c^{(0)}}\frac{C_\Sigma}{\Cso}\ach\left(\sqrt{\frac{E_c}{2T}\frac{\Delta C_g}{C_\Sigma}}\right)-
\\
\frac{e}{2}\frac{\Delta C_g}{\Cso}\sqrt{1-\frac{2T}{E_c^{(0)}}\frac{C_\Sigma}{\Delta C_g}},
\end{multline}
and
\begin{equation}
A_2=\frac{E_c^{(0)}}{eT}\frac{C_\Sigma}{\Cso}\sqrt{1-\frac{2T}{E_c^{(0)}}\frac{C_\Sigma}{\Delta C_g}}.
\end{equation}
It follows from Eqs.~\eqref{sce6} and~\eqref{eqGn} that the
conductance derivative in $\delta Q$ diverges as  $1/\sqrt{x}$
near its maximum value.

\subsubsection{The peak form at the bifurcation point~\label{secanalytical}}

To find the conductance peak at the critical value of parameter $\Delta C_g$ we
expand the hyperbolic tangents in Eq.~\eqref{sce2} up to the third order. As a result we obtain
\begin{multline}
\delta Q'\left(1-\frac{\Delta C_g}{\Cso}\left(\frac{E_c^{(0)}}{2T}-1\right)\right)+\frac{e \Delta C_g}{6 \Cso}\left(\frac{E_c^{(0)}}{T}\frac{\delta Q'}{e}\right)^3\\
=\delta Q.
\label{sce4}
\end{multline}
The linear term equals zero at the critical point. For critical polarizability of the gate-insulator we find
\begin{gather}\label{alphac}
 \Delta C_g^{(c)}=\Cso(E_c^{(0)}/2T-1)^{-1}.
\end{gather}
Also we find that
\begin{equation}
\delta Q'=\frac{eT}{E_c^{(0)}}\sqrt[3]{6\frac{\delta Q}{e}\left(\frac{E_c^{(0)}}{2T}-1\right)}.
\label{sce5}
\end{equation}
Using Eq.~\eqref{eqGn} we find that the peak maximum can be approximated with
the function $1/(1+x^{2/3})$ (here $x\propto \delta Q$), while the derivative diverges at the conductance maximum as $1/\sqrt[3]{x}$.
As follows from Fig.~\ref{figaproxn}(c) and Eq.~\eqref{sce5} this approximation for conductance works well
near its maximum value only.

\section{Single electron tunneling through slow dielectric layer~\label{SecIV}}
\begin{figure*}[t]
  \centering
  \includegraphics[width=\textwidth]{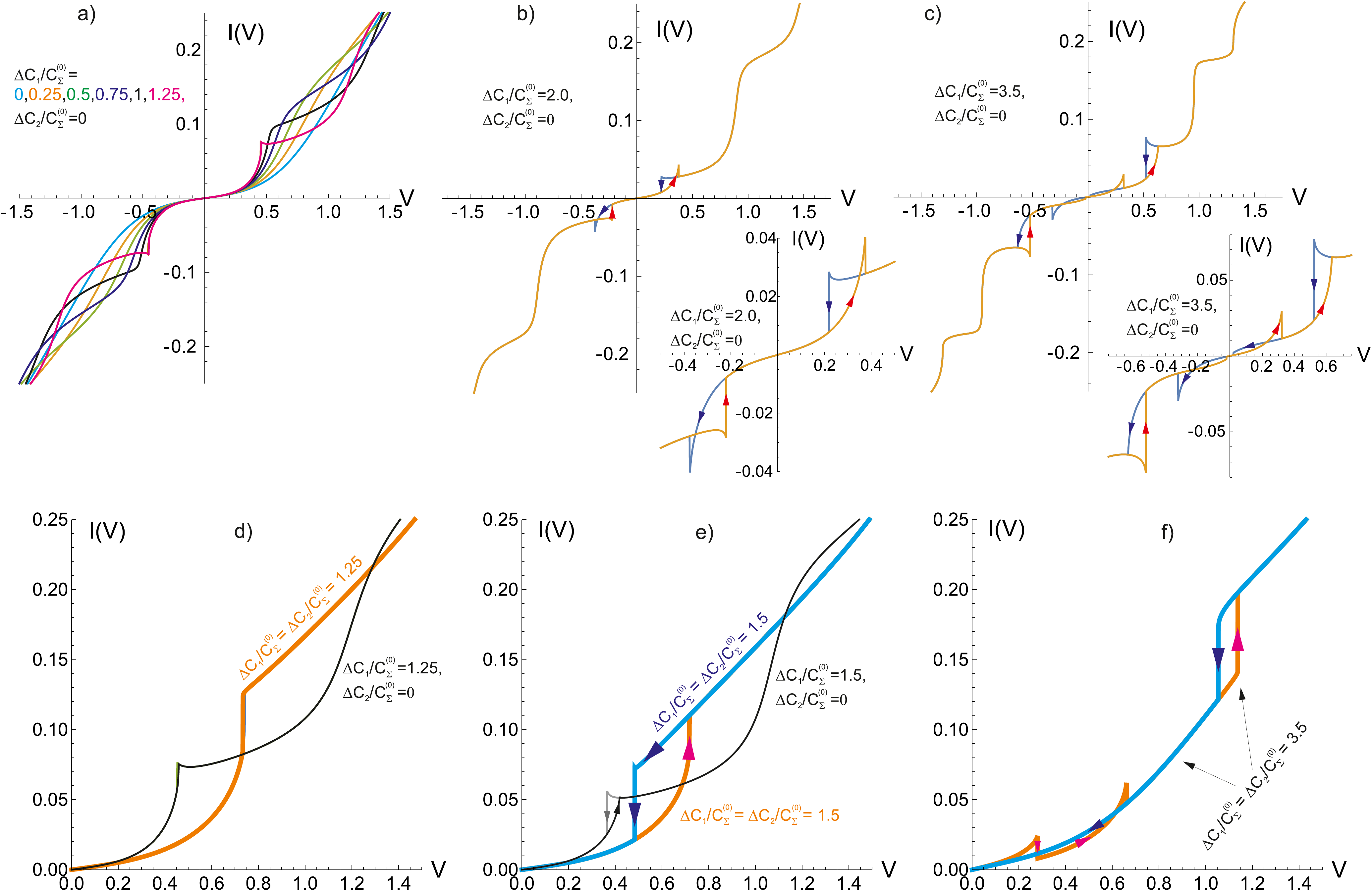}\\
  \caption{(Color online) Current-voltage characteristics, $I(V)$ of SET with zero
  gate-capacitance, $C_g=0$. Plot (a) shows $I(V)$ for $\Delta C_1/\Cso=0.0,0.25,0.5,0.75,1.0,1.25$. Smoother curves correspond to smaller $\Delta C_1$. Plot (b) corresponds to $\Delta C_1/\Cso=1.5$. The jumps in $I(V)$ in (b),(c), and (e),(f) correspond to the memory effect: the branch depends on the direction of voltage change. Plot (c) shows  $I(V)$ for $\Delta C_1/\Cso=3.5$. The $I(V)$-curve can have many hysteresis loops depending on the amount of electron charge induced on the grain by the dielectric polarization.
  Inserts in (b),(c) show the details of the hysteresis. Plot (d) shows $I(V)$ for $\Delta C_1/\Cso=1.25$, $\Delta C_2/\Cso=0$ (black curve) and  $\Delta C_1/\Cso=\Delta C_2/\Cso=1.25$ (orange curve), while plot (e) shows the graphs for $\Delta C_1/\Cso=1.5$, $\Delta C_2/\Cso=0$ (black curve) and  $\Delta C_1/\Cso=\Delta C_2/\Cso=1.5$ (orange and blue curves).
  Plot (f) shows $I(V)$ for  $\Delta C_1/\Cso=\Delta C_2/\Cso=3.5$. Parameters are:
  $T = 0.06\,E_c^{(0)}$, $C_1^{(0)}=0.6\Cso$,  $C_2^{(0)}=0.4\Cso$, and $R_i$, $i=1,2$ similar to
  Fig.~\ref{figagsmall}. The unit of voltage is $\Eco/|e|$, and  the current is normalised to $\Eco/|e|R_{\rm T1}$.}\label{figIV}
\end{figure*}
\subsection{Conductance peaks with slow dielectrics in all capacitors}

Here we consider the general case, with slow dielectric layers in all capacitors with polarizabilities
$\Delta C_1$, $\Delta C_2$ и $\Delta C_g$. Using Eq.~\eqref{eqQg} we find
\begin{gather}
Q'=Q+\Delta C_{\Sigma} \times\langle \phi \rangle(Q',V)-(\Delta C_2-\Delta C_1)\frac{V}{2},
\label{sce71}
\end{gather}
where we introduce the parameter
\begin{gather}\label{asigma}
 \Delta C_{\Sigma} =\sum_{i=1,2,g}\Delta C_i.
\end{gather}
Here we explicitly show that the functions $Q'$ and $\langle \phi \rangle$ depend on voltage $V$.
In general, this dependence results in an additional contribution to the conductance proportional to $\partial Q'/\partial V$:
\begin{multline}
G(Q,V)=\frac{\partial I^{(0)}(Q',V)}{\partial V}=\\
G^{(0)}(Q',V)+\frac{\partial I^{(0)}(Q',V)}{\partial Q'}\frac{\partial Q'}{\partial V},
\label{Geq1}
\end{multline}
where $I^{(0)}(Q,V)$ is the current in the orthodox theory, generally not
limited by the two-state approximation.
However, the current $I$ is zero for zero bias voltage for any $Q$,
therefore the last term can be omitted at $V=0$.
This explains why in two-state approximation we can calculate the conductance by
replacing $Q$ by $Q'$ in Eq.~\eqref{eqG0} of the orthodox theory.

For zero voltage, $V=0$, Eq.~\eqref{sce71} reduces to
\begin{gather}
Q'=Q+ \Delta C_{\Sigma}\times \langle \phi \rangle(Q').
\label{sce8}
\end{gather}
Then
\begin{equation}
\delta Q'\left(1+\frac{ \Delta C_{\Sigma}}{\Cso}\right)-\frac{e}{2}\frac{ \Delta C_{\Sigma}}{\Cso}\mathrm{tanh}\left(\frac{\Eco}{T}\frac{\delta Q'}{e}\right)=\delta Q.
\label{sce21}
\end{equation}
As we can see, the only distinction of the Eq.~\ref{sce21} from Eq.~\ref{sce2} is the
replacement of $\Delta C_g$ with $\Delta C_\Sigma$. It follows that for $V=0$ the SET with slow insulators in tunnel junctions
behaves qualitatively similar to the only $\Delta C_g > 0$ that was considered previously. The only difference is related to the fact that the slow dielectric in the gate capacitor renormalizes the period of the $Q$-oscillations of conductance while slow dielectrics in all other capacitors of the SET do not.

Now we can generalize our results for positive $\Delta C_g>0$ obtained earlier.
In particular, the critical polarization, where memory effect in the conductance $G(Q)$ first appears,
becomes the integral quantity, see Eq.~\eqref{asigma}, that includes properties of all the slow dielectric layers:
 \begin{gather}\label{alphacn1}
  \Delta C_{\Sigma}^{(c)}=\Cso(E_c^{(0)}/2T-1)^{-1}.
 \end{gather}
The amplitude of conductance peaks can be found using the substitution, $\Delta C_g\to\Delta C_{\Sigma}$ in Eq.~\eqref{GmaxEq}. The shape of the peaks can be obtained using the same substitution in the equations of Sec.~\ref{secanalytical}
where still $\delta Q=-(C_g^{(0)}+\Delta C_g)V_g$.

\subsection{Memory effect in current-voltage characteristics}

Above we discussed the properties of SET with slow dielectric barriers, related to the variation of the gate voltage $V_g$ at bias $V=0$.
In this subsection we instead concentrate on the current-voltage characteristic $I(V)$ of SET in the case of
electron tunneling through slow insulator in the left and
the right capacitors, see Fig.~\ref{fig_device}. We neglect the gate to simplify the situation,
putting thus $C_g=0$. Such systems have been extensively studied in experiments over the last two decades. They can exhibit coulomb blockade at room temperature(\onlinecite{Dorogi1995PRB,Nijhuis2006Small,Shinya2010APE,Klusek1999ASS}) and their ease of fabrication makes a wide range of barrier materials available for experiments.
Following Ref.~\onlinecite{Dorogi1995PRB} we consider
the current-voltage characteristic of the SET in a wide range of bias voltages.

The typical current-voltage characteristics $I(V)$ are shown in Fig.~\ref{figIV}; in  Fig.~\ref{figIV}(a)-(c) the coefficients
$\Delta C_2=0$ and $\Delta C_1$ are finite. It follows that there is a memory effect in $I(V)$ at
large enough $\Delta C_1$ and this effect depends on the direction of the bias voltage evolution.
The jumps in plots (b) correspond to the regions of hysteresis while the arrows show the evolution of voltage.
Plot (c) shows the hysteresis in $I(V)$ for $\Delta C_1/\Cso=3.5$. The current-voltage characteristics may have
many hysteresis loops, depending on the amount of electron charge that the dielectric polarization
may induce on the grain. The hysteresis in the current-voltage characteristics appears for the first time
for parameter $\Delta C_1$ being larger than $\Cso$. This is the first critical value of polarization.
For $\Delta C_1\gtrsim2\Cso$ the second hysteresis loop appears in $I(V)$. Therefore this is the second
critical value of $\Delta C_1$. For larger values of $\Delta C_1$ we expect further increase in
the number of hysteresis loops.

Two cases of current-voltage characteristics are compared in plots (e)-(d) : i) finite $\Delta C_1$ and zero $\Delta C_2$
and ii) $\Delta C_1 = \Delta C_2$. In both cases the set of critical values of $\Delta C$ is the same and for large
bias voltage the current-voltage characteristics asymptotically coincide.

Figure~\ref{figIV} shows that the current-voltage characteristics of the SET strongly depend on the direction of
bias voltage $V$. Moreover, for a given hysteresis branch
 \begin{gather}\label{IVeffect}
   I(V)\neq -I(-V)
 \end{gather}
that happens in the absence of $Q$, notably different from the result for a regular SET.

\subsection{Influence on coulomb ladder}

\begin{figure*}[t]
  \centering
  \includegraphics[width=\textwidth]{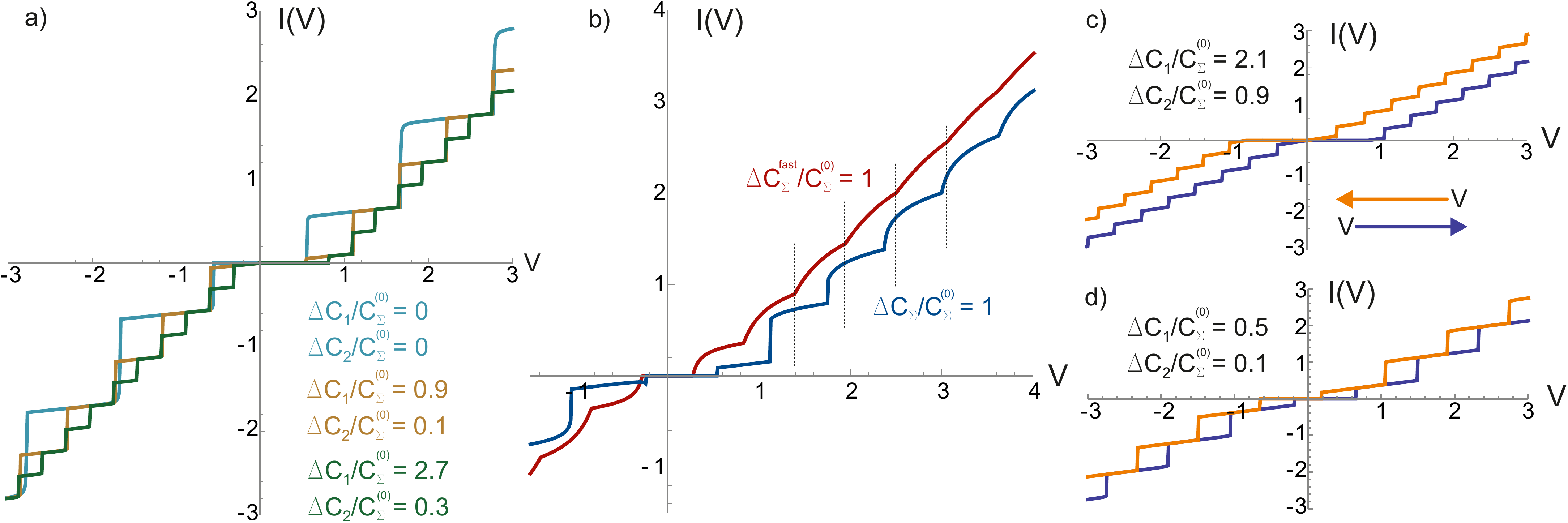}\\
  \caption{(Color online) Current-voltage characteristics, $I(V)$, of SET demonstrating the coulomb ladder at $T=0$, $C_g=0$. (a) $I(V)$ in the regime when the scaling of the coulomb ladder steps at large $V$ is the same for slow and fast dielectric response. Here $R_2/R_1=10^{-3}$, $C_1^{(0)}=0.9\Cso$, $C_2^{(0)}=0.1\Cso$. (b) The regime when the periods of coulomb ladders for slow (blue curve) and fast (red curve) dielectric of the same static polarizabilities are different. $R_2/R_1=0.25$,  $C_1^{(0)}=0.9\Cso$, $C_2^{(0)}=0.1\Cso$. (c, d) The shifts of the ladders arising from the hysteretic behavior of $I(V)$ for SET with slow dielectrics. Arrows indicate the directions of voltage change for each curve. Here $R_2/R_1=10^{-3}$,  $C_1^{(0)}=0.7\Cso$, $C_2^{(0)}=0.3\Cso$. The unit of voltage is $|e|/\Cso$, current is measured in $|e|/\Cso(R_1+R_2)$}
  \label{LadderScaleShift}
\end{figure*}

By coulomb ladder in this section we mean a step-like behavior of $I(V)$ in the regime of coulomb blockade. The coulomb ladder is often used as an indication of coulomb blockade (Ref.~\onlinecite{Dorogi1995PRB,Nijhuis2006Small,Schouteden2008SS,Oncel2005JCP,Wang2000APL}). In the following we show how the slow polarization influences the shape of the ladder. Again we take $C_g=0$ and, consider the conditions when the ladder is the most pronounced, i.e. $T=0$ and strongly asymmetric barriers $R_1\gg R_2$. At zero temperature tunneling may occur only in the direction of chemical potential drop, that is from the 1-st electrode to the 2-nd assuming $V>0$. Due to the relatively high tunneling rate through the 2-nd electrode, the number of excess electrons on the island is almost always stays at the minimum energetically allowed number $n_{min}$. $n_{min}$ can be determined as the lowest $n$ for which $\Delta F_2^{n+1\to n}<0$ is true, since $\Delta F_1^{n\to n+1}<0$ holds for any $n<0$. For a given $n_{min}$ the current can be calculated as
\begin{equation}
I=\frac{1}{e R_1}\Delta F_1^{n_{min}\to n_{min}+1},
\end{equation}
where $\Delta F_1$ is the free energy change on tunneling through the 1-st electrode. For a conventional SET the above formula leads to a ladder-shaped $I(V)$ characteristic with the step width
\begin{equation}
\Delta V_{\rm step}=|e|/C_1^{(0)},
\label{dVstep}
\end{equation}
jumps of the current between the steps
\begin{equation}
\Delta I_{\rm step}=|e|/R_1\Cso,
\end{equation}
and the $I(V)$ slope between the jumps
\begin{equation}
dI/dV=C_2^{(0)}V/\Cso R_1
\label{dIdVstep}
\end{equation}
Introducing slow dielectric into the tunnel junctions result in some new effects (for the details of
calculations see Appendix~\ref{Ap1}). At $V > |e|/\Cso$ slow polarization leads to the rescaling of the ladder that may be described by substituting the capacitances in Eqs.~\ref{dVstep}-\ref{dIdVstep} with the new values $C_i=C_i^{(0)}+\Delta C_i$, exactly as when dealing with a conventional fast dielectric (see Fig.~\ref{LadderScaleShift}(b)). But contrary to the fast dielectric, the slow one shifts the ladder, making it asymmetric and, moreover, dependent on the direction of the evolution of $V$, as illustrated at the Fig.~\ref{LadderScaleShift}(c,d). \par

Interestingly, the shift of the $I(V)$ curve in experiments is a well-known effect. It is usually accounted for by assuming the presence of some additional spurious charge $Q$, induced on the grain (as in Ref.~\onlinecite{Dorogi1995PRB,Oncel2005JCP}). However the shift that we predict is notably different at least in one aspect --- it reverses its sign with the direction of the evolution of $V$.

We stress that the described rescaling and shift of $I(V)$ takes place only under specific conditions  $V > |e|/\Cso$ and $R_1\gg R_2$. If $R_2$  are of the same order the introduction of slow dielectric may change the ladder steps in a more complex way. Such a situation is shown in Fig.~\ref{LadderScaleShift}(b) where the ladder period do not correspond to the one we would expect from the simple capacitance-renormalization consideration. If $R_1/R_2$ is even closer to unity, the slow dielectric barriers qualitatively change the current-voltage curve as was discussed in the previous section (see  Fig.~\ref{figIV}).

\section{Discussion~\label{SecDisc}}
\subsection{Requirements for dielectric materials}

Here we discuss several possible dielectric materials which can be considered as slow insulators.
At finite external electric field the localized electric charges are shifted and the dielectric material is polarized.
There are several physical processes contributing to the polarization: 1) the shift and deformation of electron-cloud, 2) the shift of
ions in the lattice, and 3) the molecular and/or macro dipole reorientation. Electrons, ions, and dipoles can form a different
polarization. The slowest polarization formation corresponds to the electrocalorical and migration (electron, ion or dipole)
mechanisms with the characteristic dispersion frequency being in the range $10^{-4}-10^{-1}$~Hz
and $10^{-3}-10^3$~Hz, respectively at temperature $T = 300K$. The electromechanical mechanism corresponds to frequencies
$10^{5}-10^{8}$~Hz, while thermal mechanism correspond to $10^{5}-10^{10}$~Hz.  The dielectrics where
thermal mechanism is the largest are promising for applications in nanostructures and can be considered
as ``slow'' dielectrics.

Dithiol self-assemble monolayers (SAMs) have a static dielectric permittivity $\epsilon(\omega =0)\sim 3$ and the characteristic relaxation frequency $\sim 10^4$~Hz.~\cite{Luo2006Chin}
These materials are good candidates for slow dielectrics. Such dielectric layers have been used in double junction SET~\cite{Dorogi1995PRB}. The hysteresis have not been observed in these experiments,  but there was a considerable discrepancy between the the values of capacitances obtained from the fit of the experimental data with the orthodox model and the ad-initio calculations.

Another promising materials to observe the hysteresis are polar crystal dielectrics e.g., BaTi$\mathrm{O_3}$ or KDP with static dielectric permittivity $\epsilon(\omega =0)\sim 10^3$ and the typical relaxation frequency $\omega_c\sim 10^6$~Hz.

\subsection{Fast capacitances}

Here we discuss the geometric capacitance $C_i^{(0)}$, $i=1,2,g$. We assumed that these capacitances has an
electrostatic origin. However, in rigorous analysis they include the high frequency dielectric permittivity $\epsilon_{\infty}$(usually between 1 and 10).
Thus in our consideration the slow polarizability $\alpha_i$ is the difference between the
low and the high frequency $\alpha_i$. As an example, for BaTi$\mathrm{O_3}$ the difference between the high and low
frequency permittivities $\epsilon$ is $\sim 10^3$. This difference is large enough.

\subsection{Critical polarization}

The effects of slow polarization are governed by the ratio of "slow" and  "fast" capacitances $\Delta C_{\Sigma}/\Cso$. If a capacitor is fully filled with a dielectric with permittivity $\epsilon(\omega)$ than $\Delta C_{\Sigma}/\Cso=(\epsilon(0)-\epsilon(\infty))/\epsilon(\infty)$. It follows from Sec.~\ref{SecII} and~\ref{SecIV} that at $\Delta C_{\Sigma}/\Cso\sim 1$ the strong influence of slow polarization may be observed, thus requiring $\epsilon(0)\gtrsim2 \,\epsilon(\infty)$.

The latter requirement become even less strict at lower temperatures. In particular, the critical value of $\epsilon(0)^{(c)}/\epsilon(\infty)$ to observe the breakdown of conductance peaks goes
to 1 as $T\to 0$ (see Eq.~\eqref{alphacn1}). For the conditions as at the Fig.~\ref{figaproxn}(e) $\epsilon(0)^{(c)}\approx 1.14 \,\epsilon(\infty)$

\begin{figure}[t]
  \centering
  \includegraphics[width=\columnwidth]{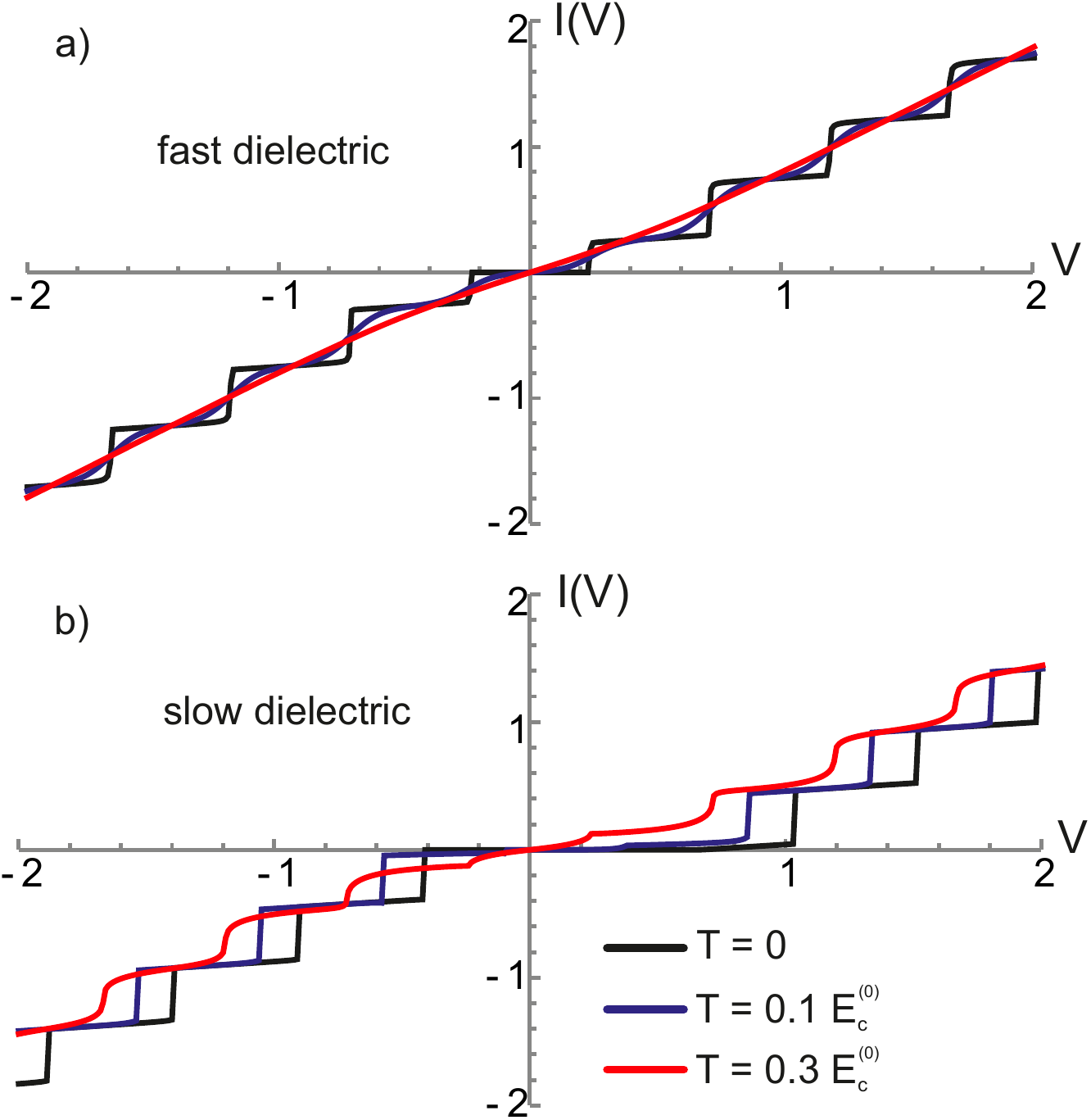}\\
  \caption{(Color online). Temperature dependence of the coulomb ladder in $I(V)$ characteristics of (a) regular SET
  and (b) SET with slow dielectrics in the tunnel barrier. In plot (b) the SET parameters are $C_1^{(0)}=0.7 \Cso$, $C_2^{(0)}=0.3 \Cso$, $\Delta C_1=1.4 \Cso$, $\Delta C_2=0$. In plot (a) $C_1=2.1\Cso$ and $C_2=0.3 \Cso$.
  For both plots $R_2/R_1=10^{-3}$ and in (b) $I(V)$ is shown for increasing voltage $V$.
  The unit of voltage is $|e|/\Cso$, the current is measured in $|e|/\Cso(R_1+R_2)$}.
\label{IVtemperature}
\end{figure}

\subsection{Temperature dependence of the coulomb-blockade effects}

A well-known consequence from the orthodox theory of SET is that in order to experimentally observe the coulomb-blockade phenomena, the temperature of the system should be lower than $E_c=e^2/2C_\Sigma$. Here the total capacitance $C_\Sigma$ includes dielectric susceptibility of the barrier media. In contrast, our numerical calculations show that if the dielectric response is sufficiently slow, only the ratio $\Eco/T$ should be taken into account when considering the blurring of the coulomb effects due to finite temperature. This must result in more pronounced blockade for a system with slow dielectric at a given temperature and electrode geometry, as illustrated in Fig.~\ref{IVtemperature}.

\section{Conclusions}

We showed that the dielectric materials at the nanoscale demonstrate new physical phenomena.
As an example we studied the single-electron transistor.
We found the memory effect in the conductance-gate voltage dependence and in the current-voltage characteristics
of the SET.  We uncovered the complex fine structure of the hysteresis-effect, where the ``large'' hysteresis
loop may include a number of ``smaller'' loops. We also found, that in order to estimate the influence of temperature on the electronic transport one should compare $T$ with $e^2 /2\Cso$ where in $\Cso$ the slow part of the dielectric function is not included.

\acknowledgments

N.C. acknowledges for the hospitality Laboratoire de Physique Théorique, Toulouse where this work was finalized and CNRS. S.~F. was supported by Russian National Foundation (Grant No. RNF 14-12-01185), N.C. by Russian Foundation of the Basic Research (grant No. 13-02-0057), the Leading Scientific Schools program No.~6170.2012.2 and SIMTECH Program, New Centure of Superconductivity: Ideas, Materials and Technologies (grant No. 246937). I.B. was supported by NSF under Cooperative Agreement Award No. EEC-1160504, NSF Award No. DMR-1158666, and the NSF PREM Award.

\appendix

\section{Calculation of the coulomb energy change on electron jumps\label{Ap2}}

Here we show, how the energy changes $\Delta U^{\pm}_n$ are calculated. If the number of electrons on the island changes from $n$ to $n\pm1$ in some process, than the electrostatic energy change is
\begin{multline}
 \Delta U^\pm_n=\int_{n}^{n\pm1}\sum_i (\phi-V_i) dq_i=\\
\int_{n}^{n\pm1}\sum_i (\phi-V_i)\left(C^{(0)}_i d\phi+S_i dP_i\right),
\end{multline}
where $q_i$ are the charges of the capacitors and $P_i$ are dielectric polarizations in barriers. For fast and slow dielectrics $P_i$ behave differently during the process of electron jump. If dielectric response is fast $P_i$ follows $\phi$ that results in capacitance renormalization.  For slow dielectric layers the polarization cannot change on the electron jump timescale and thus $dP=0$ yielding
\begin{equation}
\Delta U^\pm_n=\left. \frac{1}{2}\sum_i C^{(0)}_i  (\phi-V_i)^2\right|^{\phi(n\pm1)}_{\phi(n)}.
\label{eqDeltaUApp}
\end{equation}
$\phi(n)$ are calculated using the charge balance equation~\eqref{EqCharge0}
\begin{equation}
\phi(n)=\frac{1}{\Cso}\left[e\left(n-\sum_i P_i S_i/e\right)+\sum_i C_i^{(0)}V_i\right].
\label{eqPhiN}
\end{equation}
Here $P_i$ are constant and do not depend on $n$. By inserting Eq.~\eqref{eqPhiN} into~\eqref{eqDeltaUApp} we obtain Eq.~\eqref{eqDeltaUSlow}.
\par

\section{The shape of the coulomb ladder\label{Ap1}}

At zero temperature the tunneling rates for the electron to and from the island are
\begin{equation}
\Gamma_{1,2}^{n\to n\pm1}=\frac{1}{e^2 R_{1,2}}\left(-\Delta F_{1,2}^{n\to n\pm1}\right)\Theta\left(-\Delta F_{1,2}^{n\to n\pm1}\right),
 \label{GammaT0}
\end{equation}
where $n$ is the number of excess electrons on the island and tunneling happens through the 1-st or the 2-nd electrode. Free energy changes $\Delta F_{1,2}$ on jumps are
\begin{gather}
\Delta F_{1}^{n\to n\pm1}=\frac{e}{C^{(0)}_\Sigma}\left(\frac{e}{2}\pm\left(ne-Q'\right)\pm C^{(0)}_2 V\right)
\\
\Delta F_{2}^{n\to n\pm1}=\frac{e}{C^{(0)}_\Sigma}\left(\frac{e}{2}\pm\left(ne-Q'\right)\mp C^{(0)}_1 V \right).
\end{gather}
\par

Consider $V>0$. It follows from~\ref{GammaT0} that tunneling occurs if for some $n$ simultaneously $\Delta F_{1}^{n\to n+1}\le 0$ and $\Delta F_{2}^{n+1\to n}\le 0$ (there is no backward tunneling at $T=0$). These conditions may be combined into
\begin{equation}
 Q'/e - 1/2+C^{(0)}_1 V/e\le n\le Q'/e - 1/2-C^{(0)}_2V/e.
\end{equation}
Since the tunneling from the 1-st electrode to the island is much slower than from the island to the 2-nd electrode ($R_1\gg R_2$), the number of electrons on the island almost constantly stays at it's lowest energetically allowed value $n_{min}$. The current is, than,
\begin{equation}
I=-e\frac{\Gamma_1^{n\to n+1}\Gamma_2^{n+1\to n}}{\Gamma_1^{n\to n+1}+\Gamma_2^{n+1\to n}}\approx -e\Gamma_1^{n_{\mathrm{min}}\to n_{\mathrm{min}}+1}.
\end{equation}
\par

The rest is to calculate $n_{\mathrm{min}}$. Since we neglect $C_g$ only the charge induced by the slow polarization gives rise to $Q'$
\begin{equation}
Q'=\frac{\Delta C_\Sigma}{C_\Sigma}n_{\mathrm{min}}e+\frac{\Delta C_1 C^{(0)}_2-\Delta C_2 C^{(0)}_1}{C_\Sigma}V.
\end{equation}
$n_{min}$ can be determined from the equation
\begin{equation}
\left\lceil -\frac{1}{2}-\frac{n_{\mathrm{min}}}{1+\Delta C_\Sigma/\Cso}+\frac{C_1}{ (1+\Delta C_\Sigma/\Cso)}\frac{V}{e}\right\rceil=0,
\label{nmin}
\end{equation}
where $\lceil x \rceil$ denote the lowest integer bigger than $x$. It worth noting that the equation~\ref{nmin} predicts multiple solutions for $n_{\mathrm{min}}$ at $V$ close to the current jump points if $\Delta C_\Sigma>0$ (see Fig.~\ref{LadderScaleShift}(d)).
\par

The calculation of $I$ yields
\begin{equation}
I(V)=\frac{1}{R_1 C_\Sigma}\left(\frac{e}{2}\frac{C_\Sigma}{\Cso}+n_{\mathrm{min}}e+C_2 V\right).
\label{DeltaFnmin}
\end{equation}
The latter formula demonstrates the full renormalization of capacitances and a shift in the $I(V)$ as is illustrated at the Fig.~\ref{LadderScaleShift}(a).

\bibliography{our_bib}

\providecommand{\noopsort}[1]{}\providecommand{\singleletter}[1]{#1}%
\begin{thebibliography}{19}%
\makeatletter
\providecommand \@ifxundefined [1]{%
 \@ifx{#1\undefined}
}%
\providecommand \@ifnum [1]{%
 \ifnum #1\expandafter \@firstoftwo
 \else \expandafter \@secondoftwo
 \fi
}%
\providecommand \@ifx [1]{%
 \ifx #1\expandafter \@firstoftwo
 \else \expandafter \@secondoftwo
 \fi
}%
\providecommand \natexlab [1]{#1}%
\providecommand \enquote  [1]{``#1''}%
\providecommand \bibnamefont  [1]{#1}%
\providecommand \bibfnamefont [1]{#1}%
\providecommand \citenamefont [1]{#1}%
\providecommand \href@noop [0]{\@secondoftwo}%
\providecommand \href [0]{\begingroup \@sanitize@url \@href}%
\providecommand \@href[1]{\@@startlink{#1}\@@href}%
\providecommand \@@href[1]{\endgroup#1\@@endlink}%
\providecommand \@sanitize@url [0]{\catcode `\\12\catcode `\$12\catcode
  `\&12\catcode `\#12\catcode `\^12\catcode `\_12\catcode `\%12\relax}%
\providecommand \@@startlink[1]{}%
\providecommand \@@endlink[0]{}%
\providecommand \url  [0]{\begingroup\@sanitize@url \@url }%
\providecommand \@url [1]{\endgroup\@href {#1}{\urlprefix }}%
\providecommand \urlprefix  [0]{URL }%
\providecommand \Eprint [0]{\href }%
\providecommand \doibase [0]{http://dx.doi.org/}%
\providecommand \selectlanguage [0]{\@gobble}%
\providecommand \bibinfo  [0]{\@secondoftwo}%
\providecommand \bibfield  [0]{\@secondoftwo}%
\providecommand \translation [1]{[#1]}%
\providecommand \BibitemOpen [0]{}%
\providecommand \bibitemStop [0]{}%
\providecommand \bibitemNoStop [0]{.\EOS\space}%
\providecommand \EOS [0]{\spacefactor3000\relax}%
\providecommand \BibitemShut  [1]{\csname bibitem#1\endcsname}%
\let\auto@bib@innerbib\@empty
\bibitem [{\citenamefont {Fedorov}\ \emph
  {et~al.}(2014{\natexlab{a}})\citenamefont {Fedorov}, \citenamefont
  {Korolkov}, \citenamefont {Chtchelkatchev}, \citenamefont {Udalov},\ and\
  \citenamefont {Beloborodov}}]{RefOurPRB}%
  \BibitemOpen
  \bibfield  {author} {\bibinfo {author} {\bibfnamefont {S.~A.}\ \bibnamefont
  {Fedorov}}, \bibinfo {author} {\bibfnamefont {A.~E.}\ \bibnamefont
  {Korolkov}}, \bibinfo {author} {\bibfnamefont {N.~M.}\ \bibnamefont
  {Chtchelkatchev}}, \bibinfo {author} {\bibfnamefont {O.~G.}\ \bibnamefont
  {Udalov}}, \ and\ \bibinfo {author} {\bibfnamefont {I.~S.}\ \bibnamefont
  {Beloborodov}},\ }\href {\doibase 10.1103/PhysRevB.89.155410} {\bibfield
  {journal} {\bibinfo  {journal} {Phys. Rev. B}\ }\textbf {\bibinfo {volume}
  {89}},\ \bibinfo {pages} {155410} (\bibinfo {year}
  {2014}{\natexlab{a}})}\BibitemShut {NoStop}%
\bibitem [{\citenamefont {Fedorov}\ \emph
  {et~al.}(2014{\natexlab{b}})\citenamefont {Fedorov}, \citenamefont
  {Korolkov}, \citenamefont {Chtchelkatchev}, \citenamefont {Udalov},\ and\
  \citenamefont {Beloborodov}}]{RefOurPRBSubm}%
  \BibitemOpen
  \bibfield  {author} {\bibinfo {author} {\bibfnamefont {S.~A.}\ \bibnamefont
  {Fedorov}}, \bibinfo {author} {\bibfnamefont {A.~E.}\ \bibnamefont
  {Korolkov}}, \bibinfo {author} {\bibfnamefont {N.~M.}\ \bibnamefont
  {Chtchelkatchev}}, \bibinfo {author} {\bibfnamefont {O.~G.}\ \bibnamefont
  {Udalov}}, \ and\ \bibinfo {author} {\bibfnamefont {I.~S.}\ \bibnamefont
  {Beloborodov}},\ }\href {\doibase 10.1103/PhysRevB.90.195111} {\bibfield
  {journal} {\bibinfo  {journal} {Phys. Rev. B}\ }\textbf {\bibinfo {volume}
  {90}},\ \bibinfo {pages} {195111} (\bibinfo {year}
  {2014}{\natexlab{b}})}\BibitemShut {NoStop}%
\bibitem [{\citenamefont {Averin}\ and\ \citenamefont
  {Likharev}(1991)}]{averin1991single}%
  \BibitemOpen
  \bibfield  {author} {\bibinfo {author} {\bibfnamefont {D.}~\bibnamefont
  {Averin}}\ and\ \bibinfo {author} {\bibfnamefont {K.}~\bibnamefont
  {Likharev}},\ }\href@noop {} {\bibfield  {journal} {\bibinfo  {journal}
  {Mesoscopic phenomena in solids}\ }\textbf {\bibinfo {volume} {30}},\
  \bibinfo {pages} {173} (\bibinfo {year} {1991})}\BibitemShut {NoStop}%
\bibitem [{\citenamefont {Averin}\ \emph {et~al.}(1991)\citenamefont {Averin},
  \citenamefont {Korotkov},\ and\ \citenamefont {Likharev}}]{averin1991theory}%
  \BibitemOpen
  \bibfield  {author} {\bibinfo {author} {\bibfnamefont {D.}~\bibnamefont
  {Averin}}, \bibinfo {author} {\bibfnamefont {A.}~\bibnamefont {Korotkov}}, \
  and\ \bibinfo {author} {\bibfnamefont {K.}~\bibnamefont {Likharev}},\
  }\href@noop {} {\bibfield  {journal} {\bibinfo  {journal} {Phys. Rev. B}\
  }\textbf {\bibinfo {volume} {44}},\ \bibinfo {pages} {6199} (\bibinfo {year}
  {1991})}\BibitemShut {NoStop}%
\bibitem [{\citenamefont {Devoret}\ and\ \citenamefont
  {Grabert}(1992)}]{devoret1992single}%
  \BibitemOpen
  \bibfield  {author} {\bibinfo {author} {\bibfnamefont {M.}~\bibnamefont
  {Devoret}}\ and\ \bibinfo {author} {\bibfnamefont {H.}~\bibnamefont
  {Grabert}},\ }\href@noop {} {\emph {\bibinfo {title} {Single Charge
  Tunneling}}},\ Vol.\ \bibinfo {volume} {264}\ (\bibinfo  {publisher} {New
  York, Plenum},\ \bibinfo {year} {1992})\BibitemShut {NoStop}%
\bibitem [{\citenamefont {Wasshuber}(2001)}]{wasshuber2001computational}%
  \BibitemOpen
  \bibfield  {author} {\bibinfo {author} {\bibfnamefont {C.}~\bibnamefont
  {Wasshuber}},\ }\href@noop {} {\emph {\bibinfo {title} {Computational
  single-electronics}}}\ (\bibinfo  {publisher} {Springer},\ \bibinfo {year}
  {2001})\BibitemShut {NoStop}%
\bibitem [{\citenamefont {Arthur}(1954)}]{arthur1954hippel}%
  \BibitemOpen
  \bibfield  {author} {\bibinfo {author} {\bibfnamefont {R.}~\bibnamefont
  {Arthur}},\ }\href@noop {} {\emph {\bibinfo {title} {von Hippel, ed.:"
  Dielectric Materials and Applications}}}\ (\bibinfo  {publisher} {MIT
  Press},\ \bibinfo {year} {1954})\BibitemShut {NoStop}%
\bibitem [{\citenamefont {Thoen}\ \emph {et~al.}(1999)\citenamefont {Thoen},
  \citenamefont {Bose},\ and\ \citenamefont {Nalwa}}]{thoen1999handbook}%
  \BibitemOpen
  \bibfield  {author} {\bibinfo {author} {\bibfnamefont {J.}~\bibnamefont
  {Thoen}}, \bibinfo {author} {\bibfnamefont {T.}~\bibnamefont {Bose}}, \ and\
  \bibinfo {author} {\bibfnamefont {H.}~\bibnamefont {Nalwa}},\ }\href@noop {}
  {\emph {\bibinfo {title} {Handbook of Low and High Dielectric Constant
  Materials and Their Applications}}}\ (\bibinfo  {publisher} {Academic, San
  Diego},\ \bibinfo {year} {1999})\BibitemShut {NoStop}%
\bibitem [{\citenamefont {Ye}(2008)}]{ye2008handbook}%
  \BibitemOpen
  \bibfield  {author} {\bibinfo {author} {\bibfnamefont {Z.-G.}\ \bibnamefont
  {Ye}},\ }\href@noop {} {\emph {\bibinfo {title} {Handbook of advanced
  dielectric, piezoelectric and ferroelectric materials: Synthesis, properties
  and applications}}}\ (\bibinfo  {publisher} {Elsevier},\ \bibinfo {year}
  {2008})\BibitemShut {NoStop}%
\bibitem [{\citenamefont {Poplavko}\ \emph {et~al.}(2009)\citenamefont
  {Poplavko}, \citenamefont {Pereverseva},\ and\ \citenamefont
  {Rayevsky}}]{Poplavko2009}%
  \BibitemOpen
  \bibfield  {author} {\bibinfo {author} {\bibfnamefont {Y.~M.}\ \bibnamefont
  {Poplavko}}, \bibinfo {author} {\bibfnamefont {L.~P.}\ \bibnamefont
  {Pereverseva}}, \ and\ \bibinfo {author} {\bibfnamefont {I.~P.}\ \bibnamefont
  {Rayevsky}},\ }\href@noop {} {\emph {\bibinfo {title} {Physics of active
  dielectrics}}}\ (\bibinfo  {publisher} {Rostov: South Federal University
  Press},\ \bibinfo {year} {2009})\BibitemShut {NoStop}%
\bibitem [{\citenamefont {Dorogi}\ \emph {et~al.}(1995)\citenamefont {Dorogi},
  \citenamefont {Gomez}, \citenamefont {Osifchin}, \citenamefont {Andres},\
  and\ \citenamefont {Reifenberger}}]{Dorogi1995PRB}%
  \BibitemOpen
  \bibfield  {author} {\bibinfo {author} {\bibfnamefont {M.}~\bibnamefont
  {Dorogi}}, \bibinfo {author} {\bibfnamefont {J.}~\bibnamefont {Gomez}},
  \bibinfo {author} {\bibfnamefont {R.}~\bibnamefont {Osifchin}}, \bibinfo
  {author} {\bibfnamefont {R.~P.}\ \bibnamefont {Andres}}, \ and\ \bibinfo
  {author} {\bibfnamefont {R.}~\bibnamefont {Reifenberger}},\ }\href {\doibase
  10.1103/PhysRevB.52.9071} {\bibfield  {journal} {\bibinfo  {journal} {Phys.
  Rev. B}\ }\textbf {\bibinfo {volume} {52}},\ \bibinfo {pages} {9071}
  (\bibinfo {year} {1995})}\BibitemShut {NoStop}%
\bibitem [{\citenamefont {Luo}\ and\ \citenamefont {Xia}(2006)}]{Luo2006Chin}%
  \BibitemOpen
  \bibfield  {author} {\bibinfo {author} {\bibfnamefont {J.-l.}\ \bibnamefont
  {Luo}}\ and\ \bibinfo {author} {\bibfnamefont {C.}~\bibnamefont {Xia}},\
  }\href {\doibase http://dx.doi.org/10.1360/cjcp2006.19(6).515.4} {\bibfield
  {journal} {\bibinfo  {journal} {Chin. J. Chem. Phys.}\ }\textbf {\bibinfo
  {volume} {19}},\ \bibinfo {pages} {515} (\bibinfo {year} {2006})}\BibitemShut
  {NoStop}%
\bibitem [{\citenamefont {Vainberg}\ and\ \citenamefont
  {Trenogin}(1974)}]{VainbergTrenogin}%
  \BibitemOpen
  \bibfield  {author} {\bibinfo {author} {\bibfnamefont {M.}~\bibnamefont
  {Vainberg}}\ and\ \bibinfo {author} {\bibfnamefont {V.}~\bibnamefont
  {Trenogin}},\ }\href@noop {} {\emph {\bibinfo {title} {Theory of branching of
  solutions of non-linear equations}}}\ (\bibinfo  {publisher} {Groningen:
  Wolters-Noordhoff B. V},\ \bibinfo {year} {1974})\BibitemShut {NoStop}%
\bibitem [{\citenamefont {Nijhuis}\ \emph {et~al.}(2006)\citenamefont
  {Nijhuis}, \citenamefont {Oncel}, \citenamefont {Huskens}, \citenamefont
  {Zandvliet}, \citenamefont {Ravoo}, \citenamefont {Poelsema},\ and\
  \citenamefont {Reinhoudt}}]{Nijhuis2006Small}%
  \BibitemOpen
  \bibfield  {author} {\bibinfo {author} {\bibfnamefont {C.}~\bibnamefont
  {Nijhuis}}, \bibinfo {author} {\bibfnamefont {N.}~\bibnamefont {Oncel}},
  \bibinfo {author} {\bibfnamefont {J.}~\bibnamefont {Huskens}}, \bibinfo
  {author} {\bibfnamefont {H.~J.}\ \bibnamefont {Zandvliet}}, \bibinfo {author}
  {\bibfnamefont {B.~J.}\ \bibnamefont {Ravoo}}, \bibinfo {author}
  {\bibfnamefont {B.}~\bibnamefont {Poelsema}}, \ and\ \bibinfo {author}
  {\bibfnamefont {D.}~\bibnamefont {Reinhoudt}},\ }\href {\doibase
  10.1002/smll.200600290} {\bibfield  {journal} {\bibinfo  {journal} {Small}\
  }\textbf {\bibinfo {volume} {2}},\ \bibinfo {pages} {1422} (\bibinfo {year}
  {2006})}\BibitemShut {NoStop}%
\bibitem [{\citenamefont {Kano}\ \emph {et~al.}(2010)\citenamefont {Kano},
  \citenamefont {Azuma}, \citenamefont {Kanehara}, \citenamefont {Teranishi},\
  and\ \citenamefont {Majima}}]{Shinya2010APE}%
  \BibitemOpen
  \bibfield  {author} {\bibinfo {author} {\bibfnamefont {S.}~\bibnamefont
  {Kano}}, \bibinfo {author} {\bibfnamefont {Y.}~\bibnamefont {Azuma}},
  \bibinfo {author} {\bibfnamefont {M.}~\bibnamefont {Kanehara}}, \bibinfo
  {author} {\bibfnamefont {T.}~\bibnamefont {Teranishi}}, \ and\ \bibinfo
  {author} {\bibfnamefont {Y.}~\bibnamefont {Majima}},\ }\href
  {http://stacks.iop.org/1882-0786/3/i=10/a=105003} {\bibfield  {journal}
  {\bibinfo  {journal} {Applied Physics Express}\ }\textbf {\bibinfo {volume}
  {3}},\ \bibinfo {pages} {105003} (\bibinfo {year} {2010})}\BibitemShut
  {NoStop}%
\bibitem [{\citenamefont {Klusek}\ \emph {et~al.}(1999)\citenamefont {Klusek},
  \citenamefont {Luczak},\ and\ \citenamefont {Olejniczak}}]{Klusek1999ASS}%
  \BibitemOpen
  \bibfield  {author} {\bibinfo {author} {\bibfnamefont {Z.}~\bibnamefont
  {Klusek}}, \bibinfo {author} {\bibfnamefont {M.}~\bibnamefont {Luczak}}, \
  and\ \bibinfo {author} {\bibfnamefont {W.}~\bibnamefont {Olejniczak}},\
  }\href {\doibase http://dx.doi.org/10.1016/S0169-4332(99)00281-0} {\bibfield
  {journal} {\bibinfo  {journal} {Applied Surface Science}\ }\textbf {\bibinfo
  {volume} {151}},\ \bibinfo {pages} {262 } (\bibinfo {year}
  {1999})}\BibitemShut {NoStop}%
\bibitem [{\citenamefont {Schouteden}\ \emph {et~al.}(2008)\citenamefont
  {Schouteden}, \citenamefont {Vandamme}, \citenamefont {Janssens},
  \citenamefont {Lievens},\ and\ \citenamefont
  {Haesendonck}}]{Schouteden2008SS}%
  \BibitemOpen
  \bibfield  {author} {\bibinfo {author} {\bibfnamefont {K.}~\bibnamefont
  {Schouteden}}, \bibinfo {author} {\bibfnamefont {N.}~\bibnamefont
  {Vandamme}}, \bibinfo {author} {\bibfnamefont {E.}~\bibnamefont {Janssens}},
  \bibinfo {author} {\bibfnamefont {P.}~\bibnamefont {Lievens}}, \ and\
  \bibinfo {author} {\bibfnamefont {C.~V.}\ \bibnamefont {Haesendonck}},\
  }\href {\doibase http://dx.doi.org/10.1016/j.susc.2007.11.006} {\bibfield
  {journal} {\bibinfo  {journal} {Surface Science}\ }\textbf {\bibinfo {volume}
  {602}},\ \bibinfo {pages} {552 } (\bibinfo {year} {2008})}\BibitemShut
  {NoStop}%
\bibitem [{\citenamefont {Oncel}\ \emph {et~al.}(2005)\citenamefont {Oncel},
  \citenamefont {Hallback}, \citenamefont {Zandvliet}, \citenamefont {Speets},
  \citenamefont {Ravoo}, \citenamefont {Reinhoudt},\ and\ \citenamefont
  {Poelsema}}]{Oncel2005JCP}%
  \BibitemOpen
  \bibfield  {author} {\bibinfo {author} {\bibfnamefont {N.}~\bibnamefont
  {Oncel}}, \bibinfo {author} {\bibfnamefont {A.-S.}\ \bibnamefont {Hallback}},
  \bibinfo {author} {\bibfnamefont {H.~J.~W.}\ \bibnamefont {Zandvliet}},
  \bibinfo {author} {\bibfnamefont {E.~A.}\ \bibnamefont {Speets}}, \bibinfo
  {author} {\bibfnamefont {B.~J.}\ \bibnamefont {Ravoo}}, \bibinfo {author}
  {\bibfnamefont {D.~N.}\ \bibnamefont {Reinhoudt}}, \ and\ \bibinfo {author}
  {\bibfnamefont {B.}~\bibnamefont {Poelsema}},\ }\href {\doibase
  10.1063/1.1996567} {\bibfield  {journal} {\bibinfo  {journal} {The Journal of
  chemical physics}\ }\textbf {\bibinfo {volume} {123}},\ \bibinfo {pages}
  {044703} (\bibinfo {year} {2005})}\BibitemShut {NoStop}%
\bibitem [{\citenamefont {Wang}\ \emph {et~al.}(2000)\citenamefont {Wang},
  \citenamefont {Xiao}, \citenamefont {Huang}, \citenamefont {Sheng},\ and\
  \citenamefont {Hou}}]{Wang2000APL}%
  \BibitemOpen
  \bibfield  {author} {\bibinfo {author} {\bibfnamefont {B.}~\bibnamefont
  {Wang}}, \bibinfo {author} {\bibfnamefont {X.}~\bibnamefont {Xiao}}, \bibinfo
  {author} {\bibfnamefont {X.}~\bibnamefont {Huang}}, \bibinfo {author}
  {\bibfnamefont {P.}~\bibnamefont {Sheng}}, \ and\ \bibinfo {author}
  {\bibfnamefont {J.~G.}\ \bibnamefont {Hou}},\ }\href@noop {} {\bibfield
  {journal} {\bibinfo  {journal} {Applied Physics Letters}\ }\textbf {\bibinfo
  {volume} {77}} (\bibinfo {year} {2000})}\BibitemShut {NoStop}%
\end{thebibliography}%
\end{document}